\documentclass[showpacs,twocolumn,superscriptaddress]{revtex4}

\usepackage{doi}
\usepackage{hyperref}
\hypersetup{
  colorlinks=true,        
  linkcolor=blue,         
  citecolor=cyan,         
}

\usepackage{graphicx}
\usepackage{dcolumn}
\usepackage{bm}
\usepackage{color}
\usepackage{enumitem}
\usepackage{amsmath}
\usepackage{amssymb}

\begin{document}

\title{Shadow and quasinormal modes of the Kerr-Newman-Kiselev-Letelier black hole}

\author{Farruh Atamurotov}
\email{atamurotov@yahoo.com}

\affiliation{Inha University in Tashkent, Ziyolilar 9, Tashkent 100170, Uzbekistan}
\affiliation{Akfa University, Milliy Bog' Street 264, Tashkent 111221, Uzbekistan}
\affiliation{National University of Uzbekistan, Tashkent 100174, Uzbekistan} 
\affiliation{Tashkent State Technical University, Tashkent 100095, Uzbekistan}

\author{Ibrar Hussain}
\email{ibrar.hussain@seecs.nust.edu.pk}
\affiliation{School of Electrical Engineering and Computer Science,
National University of Sciences and Technology, H-12, Islamabad, Pakistan}

\author{G. Mustafa}
\email{gmustafa3828@gmail.com}
\affiliation{Department of Physics, Zhejiang Normal University,
Jinhua 321004, China}

\author{Kimet Jusufi}
\email{kimet.jusufi@unite.edu.mk}
\affiliation{Physics Department, State University of Tetovo, Ilinden Street nn, 
1200,
Tetovo, North Macedonia}


\begin{abstract}
We investigate the null geodesics and the shadow cast by the Kerr-Newman-Kiselev-Letelier (KNKL) black hole for the equation of state parameter $\omega_q=-2/3$ and for different values of the spacetime parameters, including the quintessence parameter $\gamma$, the cloud of string (CS) parameter $b$, the spin parameter $a$ and the charge $Q$ of the black hole. We notice that for the increasing values of the parameters $\gamma$ 
and $b$ the size of the shadow of the KNKL black hole increases and consequently the strength of the gravitational field of the black hole increases. On the other hand with increase in the charge $Q$ of the black hole the size of the shadow of the black hole decreases. Further with the increase in the values of the spin parameter $a$ of the KNKL black hole, we see that the distortion of the shadow of the black hole becomes more prominent. Moreover we use the data released by the Event Horizon Telescope (EHT) collaboration, to restrict the parameters $b$ and $\gamma$ for the KNKL black hole, using the shadow cast by the KNKL black hole. To this end, we also explore the relation between the typical shadow radius and the equatorial and polar quasinormal mods (QNMs) for the KNKL black hole and extend this correspondence to non-asymptotically flat spacetimes. We also study the emission energy rate from the KNKL black hole for the various spacetime parameters, and observe that it increases for the increasing values of both the parameters $\gamma$ and $b$ for fixed charge-to-mass and spin-to-mass ratios of the KNKL black hole. Finally, we investigate the effects of plasma on the photon motion, size and shape of the shadow cast by the KNKL black hole. While keeping the spacetime parameters fixed, we notice that with increase in the strength of the plasma medium the size of the shadow of the KNKL black hole decreases and therefore the intensity of the gravitational field of the KNKL black hole decreases in the presence of plasma. 
\end{abstract}
\maketitle

\section{Introduction}

To test gravity in the strong field regime, black holes can provide a laboratory for this purpose. Particle dynamics in the vicinity of black holes can give new insights about the structure of spacetime. Black hole shadow is a consequence of one of the predictions of the General Relativity (GR) about the bending of light by a gravitational source. To study the shadow of a black hole, the analysis of null geodesics in such spacetimes becomes important. Photons while moving in the close vicinity of black holes follow circular orbits, also known as light rings. Due to the light rings, the central black hole appears as a dark disc in the sky and known as the black hole shadow. The idea about the observation of the black hole shadow was first presented by Falcke et al. \cite{1}. The two important parameters of an astrophysical black hole are assumed to be its mass and spin. The issue of the estimation of the masses of black holes has already been resolved and black holes are now classified according to their masses in four types as stellar, intermediate, supermassive, and miniature \cite{2}. The issue of the measurement of the spin of a rotating black hole is still underway and the study of the black hole shadows is assumed to be helpful in estimating the spin of rotating black holes \cite{3}. The recent observation of the gravitational waves from a system of two merging black holes \cite{4}, and the image of the shadow of the central supermassive black holes of Messier 87 (M87*) galaxy \cite{5}, and the Milky way galaxy \cite{Akiyama2022sgr}, have further developed interest in the study of black hole spacetimes.\\  

The null geodesic of the Kerr-Newman black hole is properly studied in~\cite{Stuchlik1981,Stuchlik2000,De2000}. The investigation of null geodesics and the associated optical properties of black holes, such as black hole shadows, have been remain an active topics of research in rotating black hole spacetimes \cite{7,8,9}. The shadow cast by the Schwarzschild black hole was investigated by Synge in 1966 \cite{10}. After the work of Synge, Luminet has given a mechanism for calculating the angular radius of the shadow of a black hole \cite{11}. The shadow cast by a non-rotating black hole, like the Schwarzschild black hole, is given by a standard circle. In the case of rotating black holes, the black hole shadow elongates along the axis of rotation of the spacetime due to its dragging effects. Hioki and
Maeda suggested two observables for the characteristic points of the boundary of the shadow cast by the Kerr black hole \cite{12}. One of these observables gives description of the size of the black hole shadow, and the other one gives the deformation or deviation of the shape of the shadow from a perfect circle. These observables about the description of the size and shape of the shadow of rotating black holes have been analysed in a variety of rotating black holes \cite{13,14,15,16,Abdujabbarov2015new,Cunha2018,bambi2019,Afrin21a,Perlick2022report,Afrin2022H,Okyay2022,Amarilla2012}. Other similar works can be found in Ref.~\cite{Tsukamoto2018,Atamurotov2016a,Abdujabbarov13aa,Abdujabbarov2015b,Atamurotov22a,Ali2020izzet,Zakharov2014,Cunha2015,Ali2018,Banerjee2020,Vagnozzi2020,Eslam2020,Javier2020,Zeng2020,Jusufi2021Yang,Jafarzade21a,Junior2020,Junior2021,Dilmurod2022,Konoplya2021,Pantig2022,Fard2022,Roy2022,Ayzenberg2022,Javlon2022shadow,Guo2021a,Atamurotov2013b}.\\ 

The quasinormal modes (QNMs) of black hole spacetimes are related to the solutions of the perturbed equations subject to the boundary conditions appropriate for purely outgoing gravitational or electromagnetic waves at infinity, and purely in going such waves at the black hole horizon \cite{qnm1}. The emission of the gravitational radiations by black holes do not allow normal mode of oscillations \cite{qnm2}, and therefor the emitted frequencies of oscillation become quasinormal (complex). The real part of these complex frequencies corresponds to the actual frequency of the oscillation and the imaginary part represents the damping of the frequency \cite{qnm3}. Using the notion of geometric-optics and the conserved quantities along geodesics in black hole spacetimes, an equation has been derived which links the equatorial and polar QNMs to the radius of the shadow cast by a black hole \cite{Jusufi:2020dhz}. The equation relating the QNMs to the shadow radius has been applied to investigate the radius of the shadow cast by some particular black holes, including the Kerr black hole, the Kerr-Newman black hole and the five dimensional Myers-Perry black hole \cite{Jusufi:2020dhz}. In the present work we modify the equation derived in \cite{Jusufi:2020dhz}, for non-asymptotically flat spacetimes and apply it to the KNKL black hole.\\ 

In the year of 1974, Hawking used quantum field theory in the curved background and gave the idea that a black hole can radiate, and this phenomenon is known as the Hawking radiation \cite{17}. Due to the Hawking radiation, which reduces the mass and the rotational energy of a black hole, it is possible for a black hole to evaporate in a certain period of time. The evaporation of a black hole can be thought of in terms of the rate at which the energy is emitted by the black hole. The energy emission rate of black holes therefore is a topic of interest. The rate at which the energy emitted by a black hole is directly linked with the radius of the shadow cast by the black hole and has been investigated for different black holes in the literature \cite{18,19,20,20a}. The effects of different black hole spacetimes parameters have been analysed on the energy emission rate of black holes. For the Einstein-Maxwell-Dilaton-Axion black hole, Wei and Liu have investigated the energy emission rate which decreases with the Dilaton parameter for fixed values of the spin parameter of the black hole \cite{21}. Papnoi et al. have reported that the rate of the emission of energy in the five dimensional Myers-Perry black hole decreases as the value of the spin parameter of the black hole increases \cite{22}. For some regular rotating black holes, namely, Ayon-Beato-Garcıa, Hayward, and Bardeen black holes, Abdujabbarov et al. have analysed the emission energy rate \cite{23}. For these regular black holes they have observed a decrease in the emission energy rate with the electric as well as the magnetic charge of the black holes. In a recent work the emission energy rate for a charged rotating black hole surrounded by perfect fluid dark matter has been investigated by Atamurotov et al. \cite{24}. They have shown that for the increasing values of the perfect fluid dark matter parameter and charge of the black hole, this emission rate decreases.\\     

Another interesting astrophysical scenario is that, photons mostly go through a plasma medium in the surrounding of black holes. The plasma medium can have very important impact on the angular positions of an equivalent image of the black hole shadow and therefore, give various wavelengths in observations. This is one of the reason people consider the plasma medium in the analysis of the shadow cast by black holes. To explore the formation of jets emitted by black holes a simplified model of plasma has been considered in black hole environments \cite{PL1}. Kogan and Tsupko have discussed the effects of the nonuniform plasma medium on the gravitational deflection angle of photon by the Schwarzschild black hole \cite{25}. The shadows cast by the Schwarzschild and Kerr black holes have been considered in the plasma medium by utilising the radial power-law density \cite{26}, where it has been noticed that for both the black holes the size and also the shape of the shadows get effected due to the presence of the plasma. The dependence of the plasma on the gravitational deflection angle of photon and the shadow cast by the charged-Kerr black hole have been studied in \cite{27}.  Other recent interesting works related to the study of the effects of plasma on the shadows cast by black holes in the GR and other modified theories of gravity can be found in the literature  \cite{28,29,30,31,32,33,Farruh2021}.\\ 

From the mathematical point of view black holes are the singular solutions of Einstein's fields equations of GR. Black hole solutions have also been found in other alternative theories of gravity \cite{34,35,36}. In the GR first such solution was obtained by Schwrazschild for a point gravitating source \cite{37}. The Schwarzschild solution was later generalized for an axially symmetric rotating gravitating source by Kerr \cite{38}. The Kerr black hole solution is assumed to be the most suitable solution describing an astrophysical black hole \cite{39}, and therefore is widely studied in different contexts \cite{40,41,42,43,44}. The Kerr black hole solution has been further generalized in the presence of charge \cite{45}, dark energy \cite{46}, CS \cite{47} and other fields \cite{48,49}, both in the asymptotically flat and asymptotically de-sitter/anti-de-sitter backgrounds. The current accelerated expression of the Universe can be possibly explained with the existence of a repulsive gravitational force for which the dark energy is a potential candidate. There are different notions of the dark energy introduced in the literature. The quintessential dark energy for which the equation of state parameter $\omega_{q}$ is given as $(-1;-1/3)$, 
has been considered widely in the different gravitational theories \cite{50,51,52}. In astrophysical situations, the dark energy and in particular the quintessence can produces gravitational effects in the black hole spacetimes on the deflection angle of light that coming from distant stars, and should be taken into consideration in such scenario. To look at the imprints of the quintessence on the black hole spacetimes, a black hole solution in the presence of the quintessence has been obtained by Kiselev for the first time \cite{53}. A rotating Kislev solution with charge has also been presented \cite{47}. To quantize gravity, the string theory is an attempt in this direction. According to the string theory the fundamental ingredients of the Universe are one-dimensional strings instead of point particles. The first static black hole solution in the background of CS has been obtained by Letelier in 1979 \cite{54}. These CS are assumed to be formed in the very early stages of the structure formation in the Universe due to the symmetry breaking \cite{55}. Toledo and Bezerra have combined the ideas of Kiselev and Letelier about the black hole solutions, and have obtained the counterpart of the Kerr-Newman solution in the presence of both the quintessential dark energy and CS \cite{47}. Recently, some interesting work has been done for black holes in the presence of dark energy and CS \cite{Fathi1,Fathi2,Mustafa2021}. In the current work we are going to explore the photon motion and the shadow cast by the Kerr-Newman-Kislev-Letelier (KNKL) black hole. We show that how the photon sphere and shadow of the KNKL black hole varies with the quintessence parameter, the CS parameter and the black hole parameters, namely, charge and spin of the black hole. Besides, we also examine the effects of a plasma medium on the photon motion and the shadow cast by the KNKL black hole. The shadow size increases with the quintessence and CS parameter while it decreases with the charge of the KNKL black hole. Further we investigate the rate of energy emission by the KNKL black hole which increase with the quintessence and CS parameters for fixed values of the spin and charge of the KNKL black hole.\\     

The paper is organized as follows: In Section \ref{Sec:metricKNK} we briefly review the KNKL black hole and it horizon structure. In the same Section we study the null geodesics in the KNKL black hole spacetime. We investigate the shadow cast by the KNKL black hole in the Section \ref{Sec:shadow}, where we also analyse the two observables relating to the size and shape of the shadow of the KNKL black hole. Further in the same Section we calculate limits on the parameters $\gamma$ and $b$ for the KNKL black hole, from the data provided by the EHT collaboration. The rate of emission energy for the KNKL black hole is discussed in Section \ref{Sec:emission}. Section \label{Sec:QNMs} is devoted to the study of the equatorial and polar QNMs and their correspondence with the radius of the shadow cast by the KNKL black hole. In the Section \ref{Sec:plasmashadow} we look at the effects of plasma on the null geodesics and shadow of the KNKL black hole. We present conclusion and discussion in the Section \ref{Sec:Conclusions}.    

\section{Null geodesics in the Kerr-Newman-Kiselev-Letelier black hole}
\label{Sec:metricKNK}
\begin{figure*}
 \begin{center}
   \includegraphics[scale=0.65]{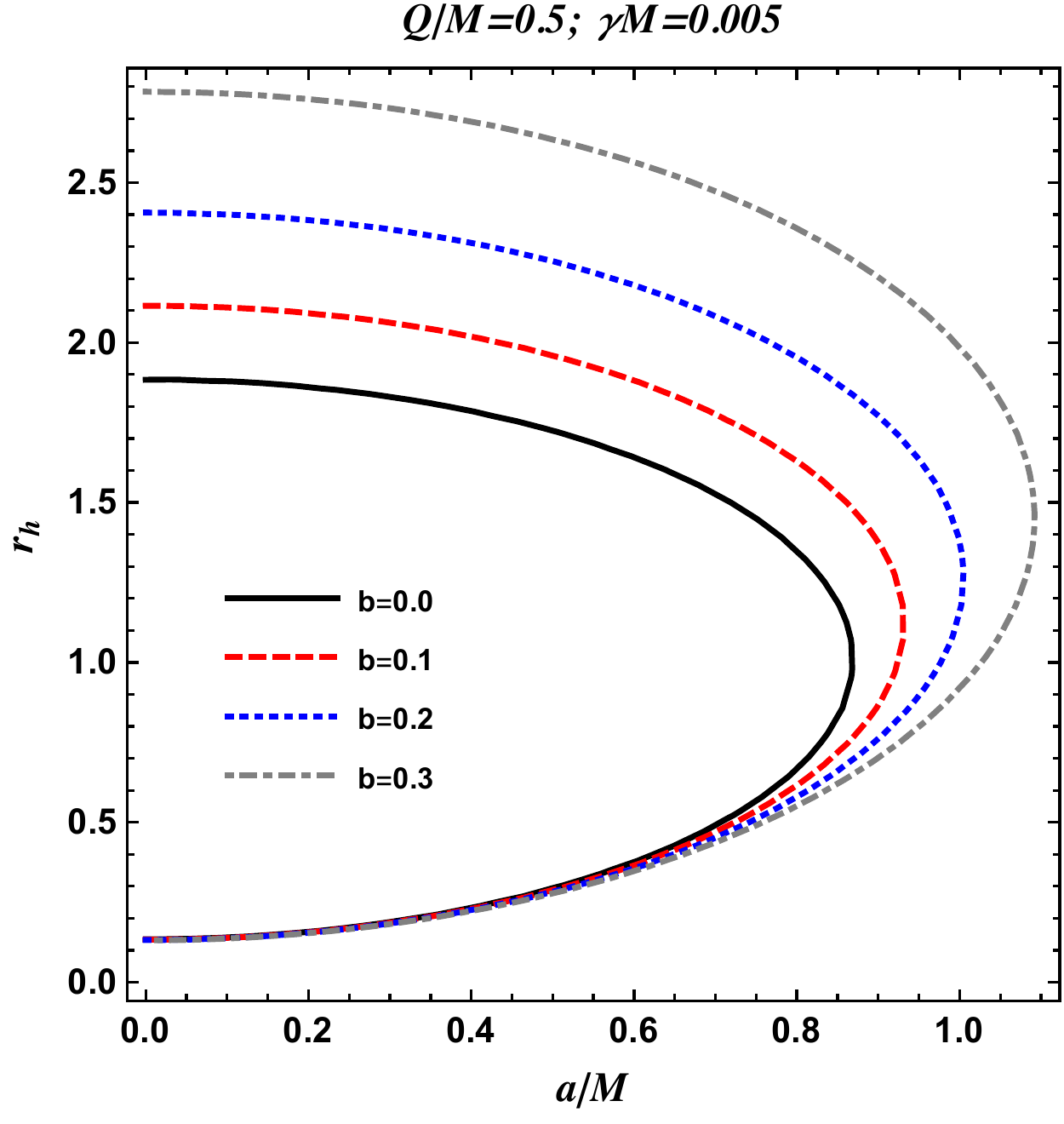}
   \includegraphics[scale=0.65]{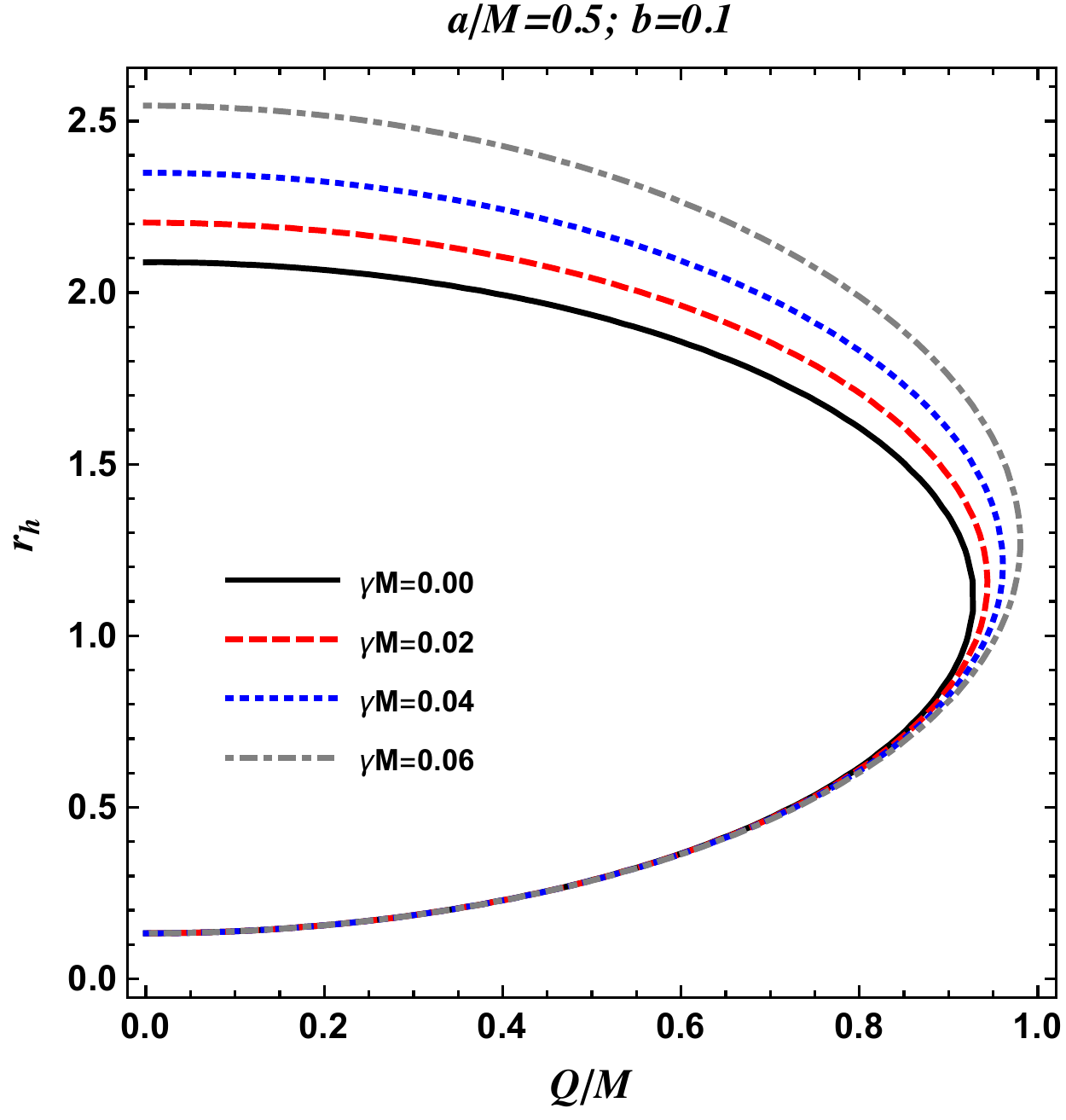}
  \end{center}
\caption{The plots show the horizons structure of the KNKL black hole for different values of the parameters.}\label{plot:horizon}
\end{figure*}

The KNKL black hole metric can be presented as \cite{47}
\begin{eqnarray}
 ds^2 &=& -\frac{\Delta}{\Sigma}(dt-a\sin^2\theta d\phi)^2+\frac{\Sigma}{\Delta}dr^2+\Sigma d\theta^2
  \nonumber \\ && +\frac{\sin^2\theta}{\Sigma}\left(a dt-(r^2+a^2)d\phi\right)^2.\label{Romet}
\end{eqnarray}
For the above metric (\ref{Romet}), the metric coefficients are
\begin{eqnarray}
 \Sigma&=&r^2+a^2\cos^2\theta,\\
 \Delta&=&(1-b)r^2+a^2+Q^2-2Mr-\gamma r^{-3\omega_{q}+1}.\label{DDm}
\end{eqnarray}

Here we discuss the effect of quintessence parameter $\gamma$, CS parameter $b$, spin parameter $a$ and charge $Q$ of the black hole on the structure of the horizons of the KNKL black hole. Note that $M$ is the mass parameter and the ADM mass of the system can be obtained by rescaling the coordinates (see, \cite{Vagnozzi:2022moj}). The horizon of the black hole spacetime (\ref{Romet}) can be obtained by setting $\Delta =0$. Solving $\Delta =0$ numerically, the behaviour of the radius of the horizon for different values of the parameters $\gamma$ and $b$ is shown in Fig. (\ref{plot:horizon}) for fixed values of the spin parameter $a$ and charge $Q$ of the KNKL black hole. From the Fig. (\ref{plot:horizon}) one may see that the radius of the horizon of the KNKL black hole increases with the increasing values of the parameters $\gamma$ and $b$, and therefore, the radius of the horizon of the Kerr black hole is smaller than the radius of the horizon of the KNKL black hole. For further discussion on the horizon structure of the KNKL black hole one may see Re. \cite{47}.

 In order to have the shape of the shadow cast by the KNKL black hole we first investigate null geodesics in the spacetime geometry of the KNKL black hole. We adopt the Hamilton-Jacobi formalism to study the null geodesic structure of the KNKL black hole spacetime, as follows \cite{PhysRev.174.1559}

\begin{equation}
 \frac{\partial S}{\partial\tau}=-\frac{1}{2}g^{\mu\nu}\frac{\partial S}{\partial x^{\mu}}\frac{\partial S}{\partial x^{\nu}},\label{sequation}
\end{equation}
where $E=g_{t\mu}\dot{x}^{\mu}$ and $L_z=g_{\phi\mu}\dot{x}^{\mu}$  correspond to the energy and momentum of the particle, $\tau$ is the affine parameter and $g^{\mu\nu}$ is the metric tensor. Here the action $S$ is separated as
\begin{equation}
 S=\frac{1}{2}m_0^2\tau-Et+L_z\phi+S_{r}(r)+S_{\theta}(\theta),\label{jaction}
\end{equation}
where $m_{0}$ is the mass of the particle and $S_r(r)$, $S_{\theta}(\theta)$ are the function of $r$ and $\theta$ only. Using the method of separation of variables and utilizing Eq. (\ref{jaction}) into the Eq. (\ref{sequation}), we obtain the following equations for the motion of photon $(m_0 = 0)$

\begin{eqnarray}
 \mathcal{R}&=&\Big[(r^2+a^2)E-aL_z\Big]^2-\Delta\Big[\mathcal{K}+(L_z-aE)^2\Big],\\
 \Theta&=&\mathcal{K}+\cos^2\theta\left(a^2 E^2-L_z^2\sin^{-2}\theta\right).
\end{eqnarray} 

We obtain the following geodesic equations in the spacetime metric of the KNKL black hole
\begin{eqnarray}
 \Sigma\frac{dt}{d\tau}&=&a(L_z-aE\sin^{2}\theta)
       \nonumber \\ &&+\frac{r^{2}+a^{2}}{\Delta}\Big(E(r^{2}+a^{2})-aL_z\Big),\label{rhot}\\
 \Sigma\frac{dr}{d\tau}&=&\pm\sqrt{\mathcal{R}},\label{Rad}\\
 \Sigma\frac{d\theta}{d\tau}&=&\pm\sqrt{\Theta},\\
 \Sigma\frac{d\phi}{d\tau}
     &=&(L_z\csc^{2}\theta-aE)+\frac{a}{\Delta}\Big(E(r^{2}+a^{2})-aL_z\Big).\label{rhophi}
\end{eqnarray}
To obtain the boundary of the shadow cast by the KNKL black hole we consider the geodesic equation involving $\dot{r}$ and can be written as 
\begin{equation}
 \left(\Sigma\frac{dr}{d\tau}\right)^2+V_{eff}=0.
\end{equation}
The effective potential in the equatorial plane $(\theta = \pi/2)$ is given as
\begin{eqnarray}\label{veff}
V_{eff}&=&\frac{\Delta ({\cal K} + (L_z - a E)^2)-((a^2 + r^2) E - a L_z)^2}{2r^4}.
\end{eqnarray}
\begin{figure*}
 \begin{center}
   \includegraphics[scale=0.7]{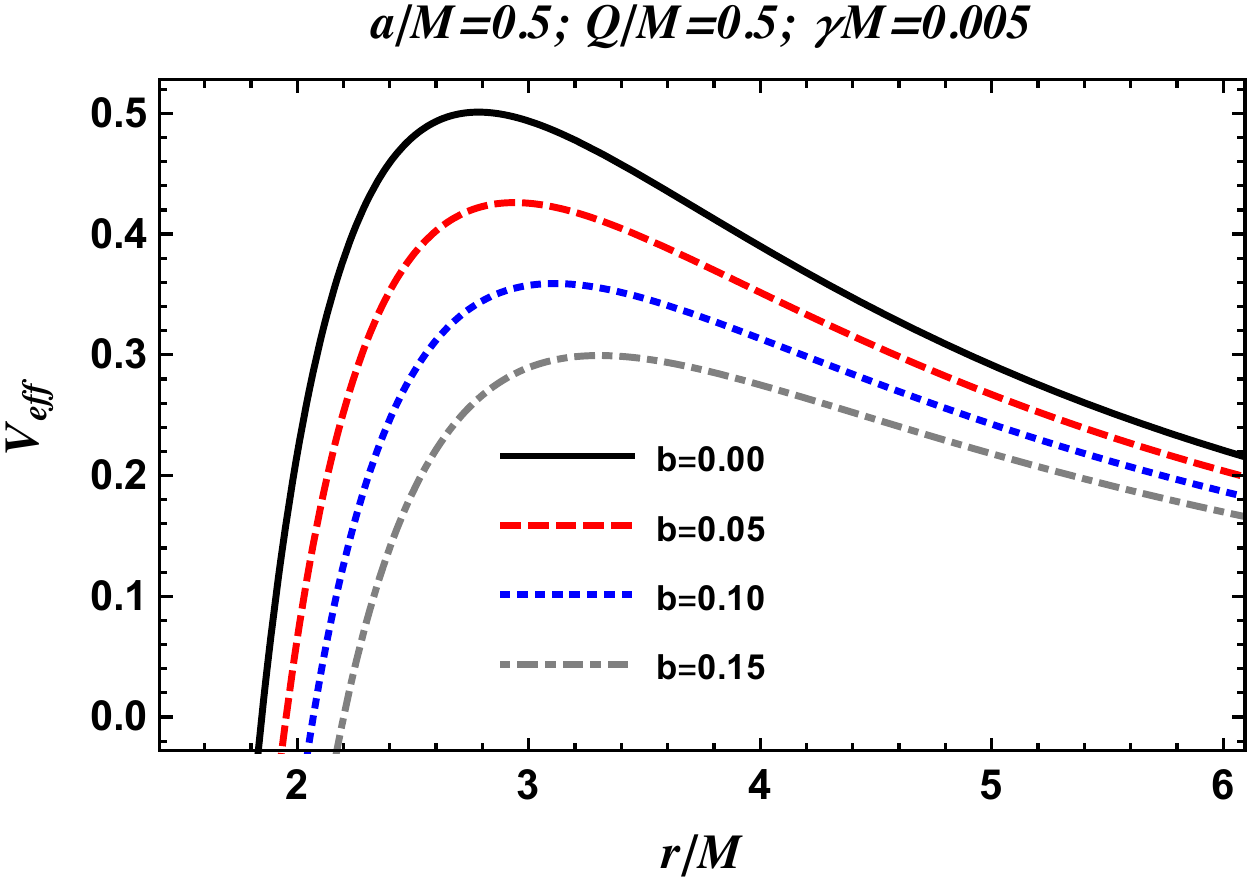}
   \includegraphics[scale=0.7]{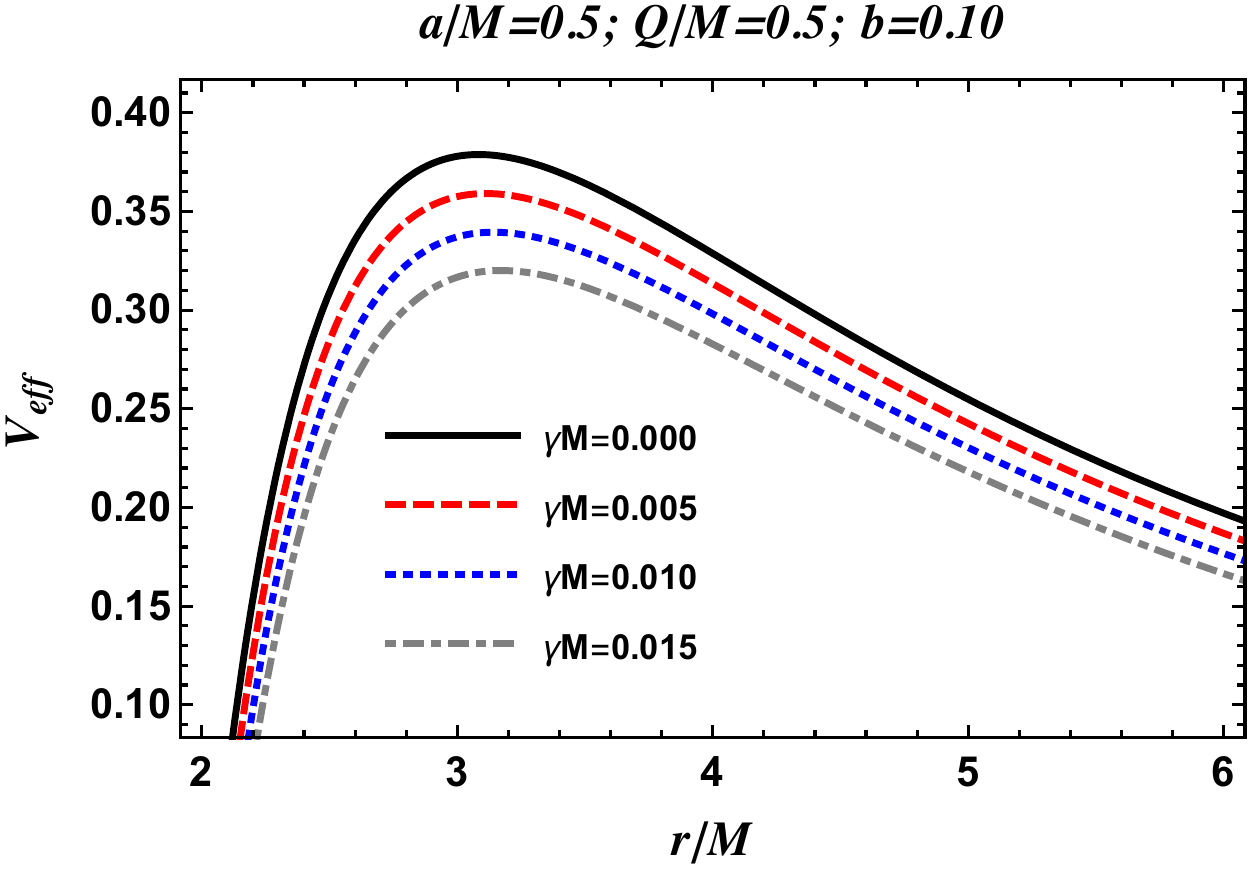}
  \end{center}
\caption{Effective potential around KNKL black hole for the massless particle.}\label{plot:effpot}
\end{figure*}

In Fig. (\ref{plot:effpot}), we show the general behaviour of the effective potential against the radial coordinate $r$ for some different values of the CS parameter $b$ and the quintessence parameter $\gamma$, for fixed values of the spin and charge of the KNKL black hole. The unstable circular orbits of photons can be seen from the graphs of the effective potential, where the maxima or peaks correspond to these orbits. From the same graphs for the effective potential, we notice that the peaks are decreasing and moving towards the right with increase in the value of the parameters $b$ and $\gamma$ for fixed values of the spin $a$ and charge $Q$ of the KNKL black hole. Therefore we observe that the unstable circular orbits are shifting away from central object with the increasing values of the parameters $b$ and $\gamma$. We further observe that the circular orbits of photons become more and more unstable as the values of the parameters $b$ and $\gamma$ goes on decreasing. from the behaviour of the effective potential seen here, we confirms the repulsive nature of the quintessential dark energy. It further indicates that the CS can also have repulsive nature in the spacetime of the KNKL black hole.

\section{The shadow cast by the Kerr-Newman-Kiselev-Letelier black hole and the observables}
\label{Sec:shadow}
Here we determine the shape of the shadow of the KNKL black hole. For this we define the following impact parameters $\xi = L_z/E$ and $\eta = \mathcal{K} /E^2$, and therefor, the $\mathcal{R}$ in terms of these new parameters can be expressed as 

\begin{eqnarray} \label{r}
 \mathcal{R}&=&\Big[(r^2+a^2)-a\xi\Big]^2-\Delta\Big[\eta+(\xi-a)^2\Big].
 \end{eqnarray} 

Photons coming towards a black hole follow three types of trajectories, i.e., falling inside the black hole, scattering away from the black hole or moving in unstable circular orbit in the vicinity of the horizon of the black hole. Out of these three types of trajectories, the unstable circular orbits near the horizon of a black hole are responsible for determining the shape of the shadow cast by the black hole. The unstable circular photon orbits can be obtained from the following conditions
\begin{figure*}
 \begin{center}
   \includegraphics[scale=0.65]{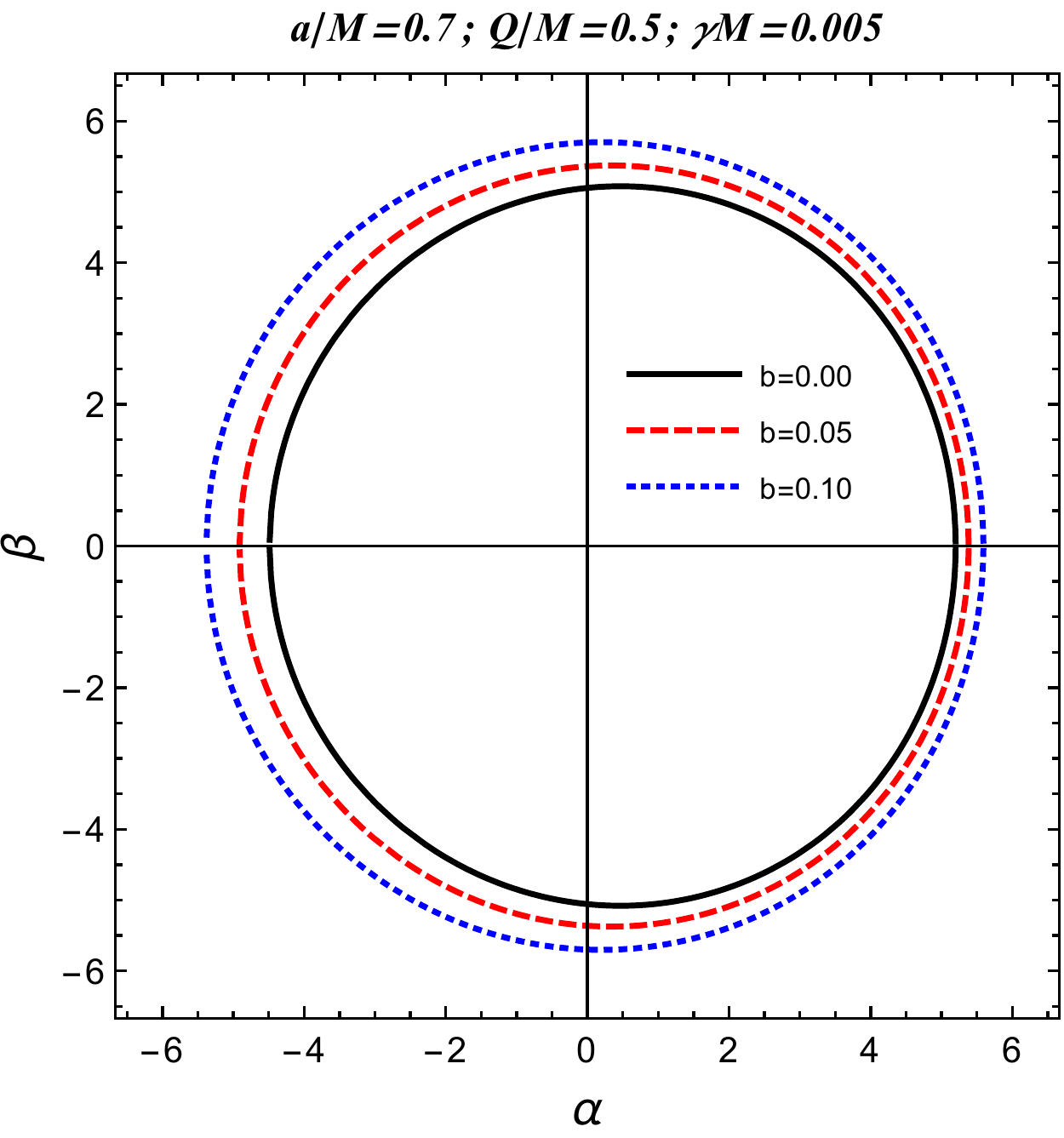}
   \includegraphics[scale=0.65]{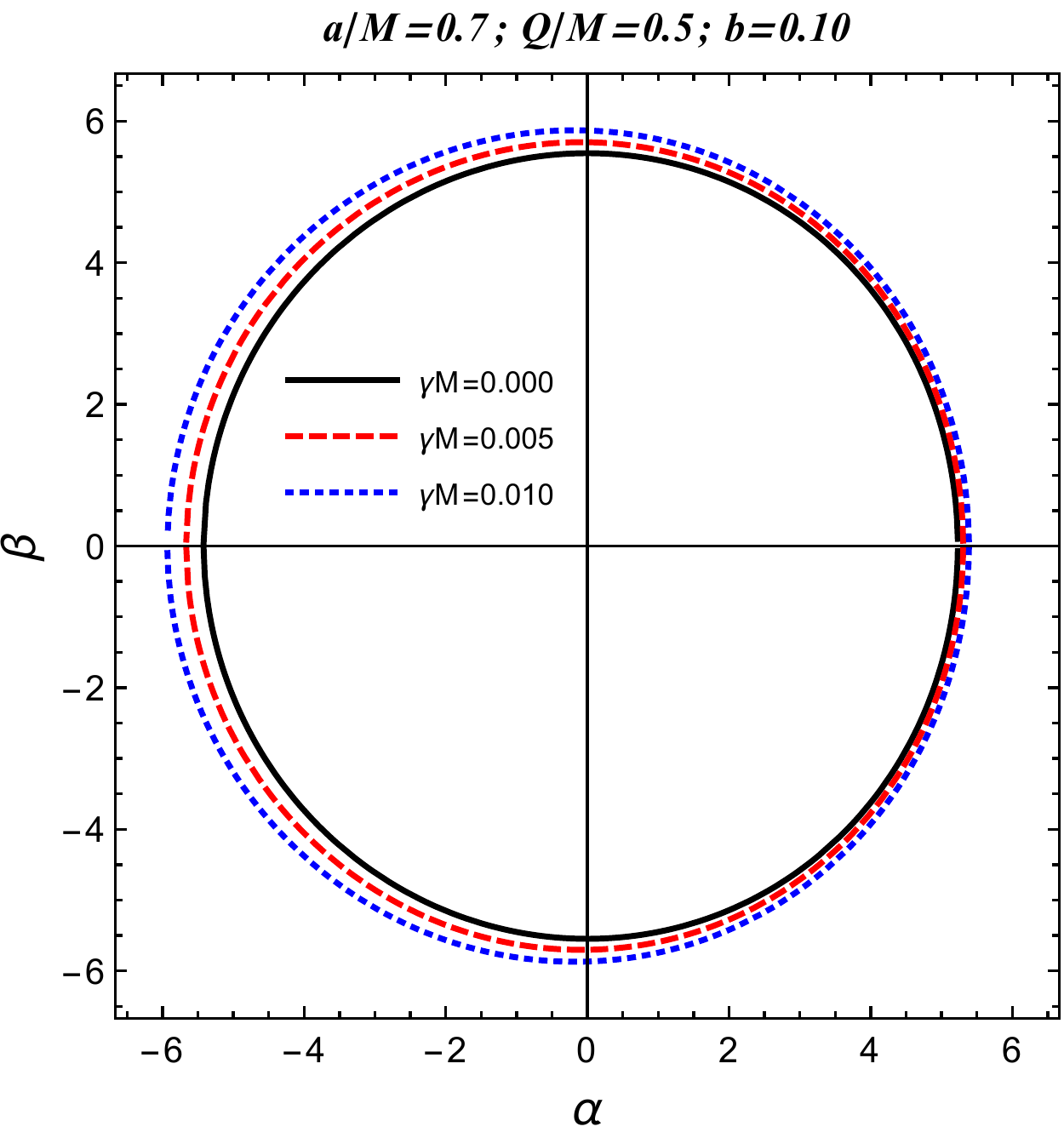}
   \includegraphics[scale=0.65]{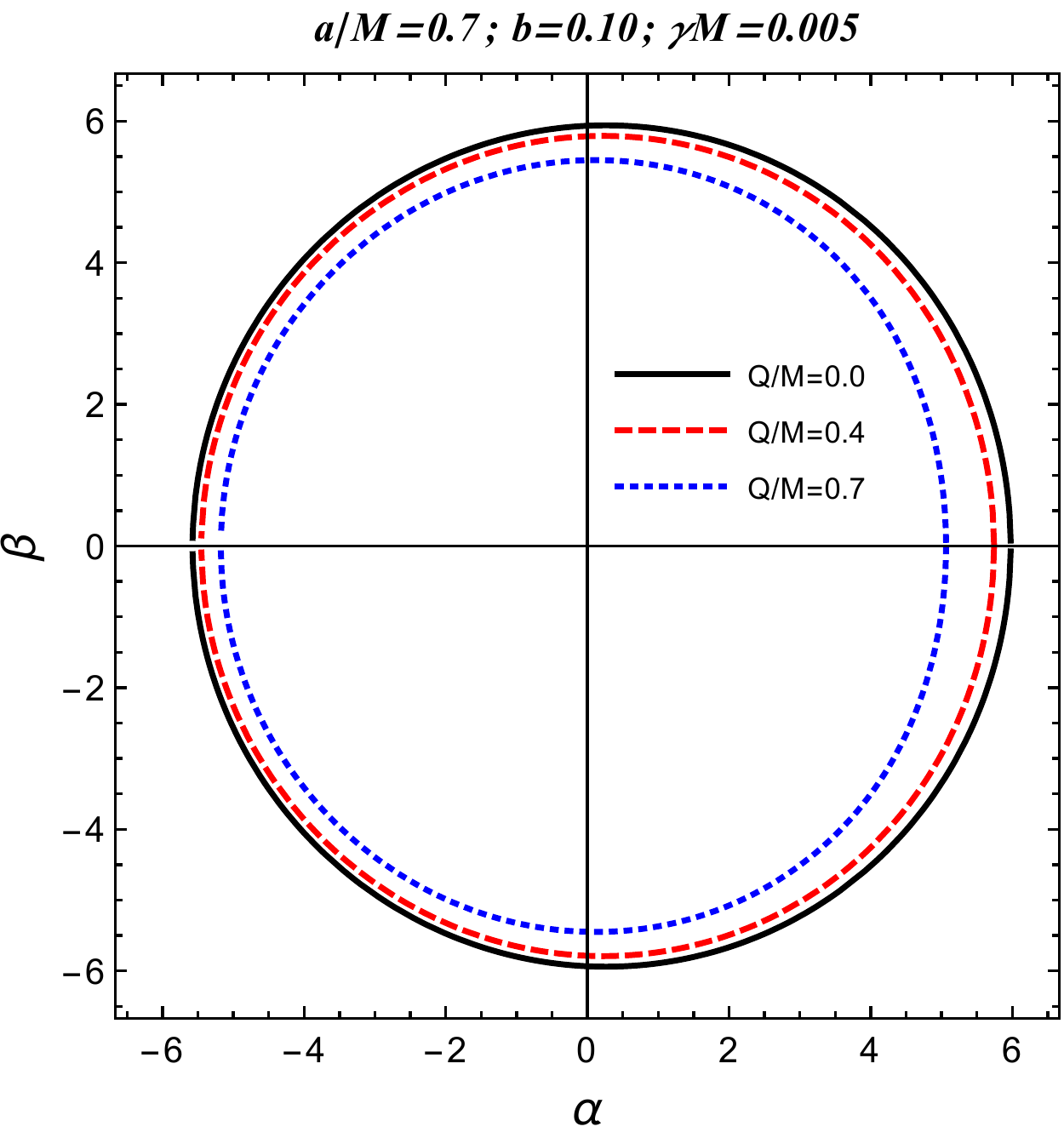}
   \includegraphics[scale=0.65]{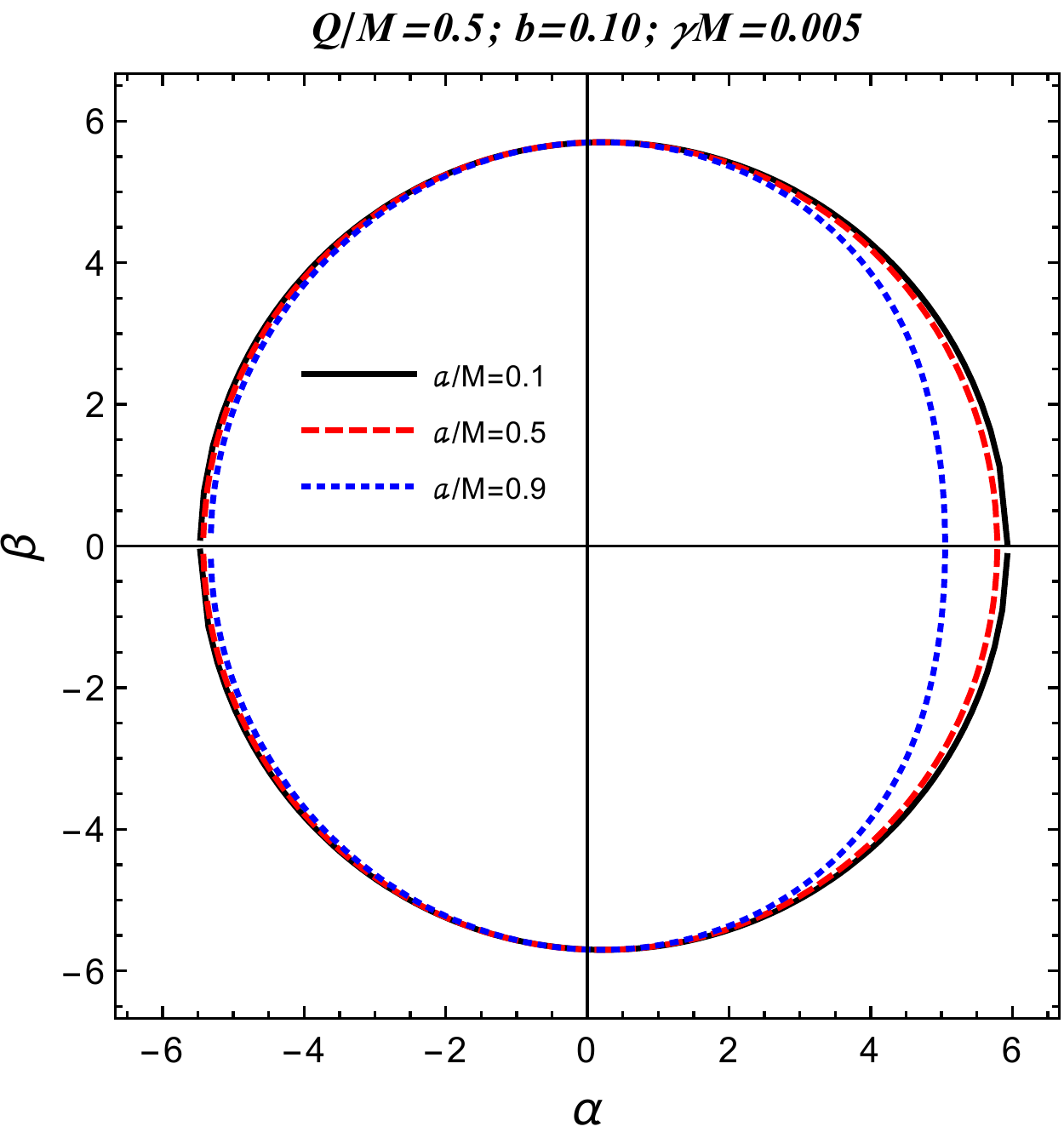}
  \end{center}
\caption{Shadow cast by the Kerr-Newman-Kiselev-Letelier black hole and  the observables. }\label{plot:shadow}
\end{figure*}

\begin{equation}\label{qqq}
\mathcal{R}(r)=0=\frac{\partial{\mathcal{R}(r)}}{\partial{r}}.
\end{equation}
The two parameters $\xi$ and $\eta$ determine the shape of the shadow cast by a black hole. Considring  Eqs. (\ref{r}) and (\ref{qqq}), for the KNKL black hole we have $\xi$ and $\eta$ as

\begin{eqnarray}
 \xi&=&\frac{\left(a^2+r^2\right) \Delta'-4 \Delta r}{a \Delta'},\label{xx1}\\
 \eta&=&\frac{r^2 \left(16 \Delta  \left(a^2-\Delta\right)-r^2 \Delta'^2+8
   \Delta  r \Delta'\right)}{a^2 \Delta'^2},\label{xx2}
\end{eqnarray}

To discuss the geometry of the shadow cast by the KNKL black hole the celestial coordinates $\alpha$ and $\beta$ are defined as \cite{Kimetnoflat2020}
\begin{eqnarray}
 \alpha&=& -r_{0} \frac{P^{(\phi)}}{P^{(t)}},\label{alpha} \\
 \beta&=&-r_{0} \frac{P^{(\theta)}}{P^{(t)}}.\label{beta}
\end{eqnarray}

where $P^{(\phi)}$, $P^{(\theta)}$, and $P^{(t)}$ are the tetrad components
of the photon momentum with respect to locally nonrotating reference frame \cite{Kimetnoflat2020}.The $r_o$ is the observed distance and it is very large but finite $r_{0}=D=8.3$ kpc for the Sgr A* or $r_{0}=D=16.8$ Mpc for the M87*.

We are determining the shadow of the KNKL black hole in the equatorial plane for which the angle of the inclination is $\theta_0 = \pi/2$, therefore,  Eqs. (\ref{alpha}) and (\ref{beta}) assumes the following form
\begin{eqnarray}
 \alpha&=&-\sqrt{-g_{tt}(r_0)}\xi\\
 \beta&=&\pm\sqrt{-g_{tt}(r_0)}\sqrt{\eta}.
\end{eqnarray}
To analyse the shape and size of the shadow cast by the KNKL black hole we plot the two celestial coordinates $\alpha$ and $\beta$ for different values of the parameters $b$ and $\gamma$ for fixed values of the spin parameter $a$ and charge $Q$ of the black hole in Fig. \ref{plot:shadow}. We observe that for increasing values of the parameters $b$ and $\gamma$ the radius of the shadow cast by the KNKL black hole increases. In the same Fig. \ref{plot:shadow} we plot the coordinates $\alpha$ and $\beta$ for different values of the spin parameter $a$ and charge $Q$ of the KNKL black hole, keeping the parameters $b$ and $\gamma$ fixed. Here we notice that the shadow radius for the KNKL black hole gets smaller as the values of the parameters $a$ and $Q$ go on increasing. Again the repulsive nature of both the quintessence and CS can be confirmed from this behaviour of the graphs for the shapes of the shadow cast by the KNKL black hole. Further, form the Fig. \ref{plot:shadow} we observe that the increasing values of the spin parameter $a$ enhances the distortion of the shadow of the KNKL black hole, for fixed values of the other spacetime parameters. Thus the shadow cast by a fast rotating black hole would be more distorted as compared to the one cast by a slowly rotating black hole. This observation may be helpful in constraining the spin of a rotating black hole. Here we see that  all the spacetime parameters have an influence on the shape and size of the shadow cast by the KNKL black hole.
          
\begin{figure*}
 \begin{center}
   \includegraphics[scale=0.5]{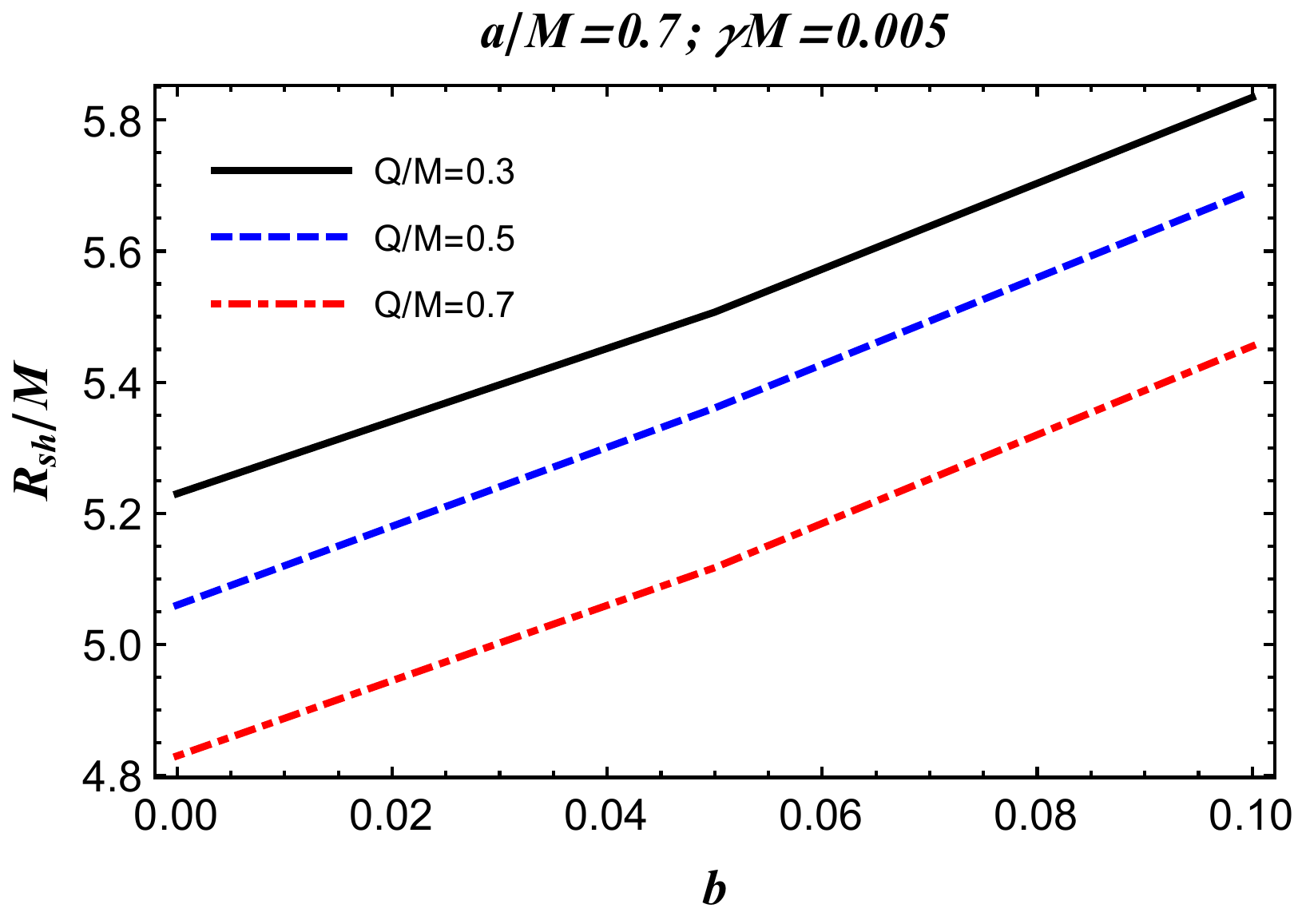}
   \includegraphics[scale=0.5]{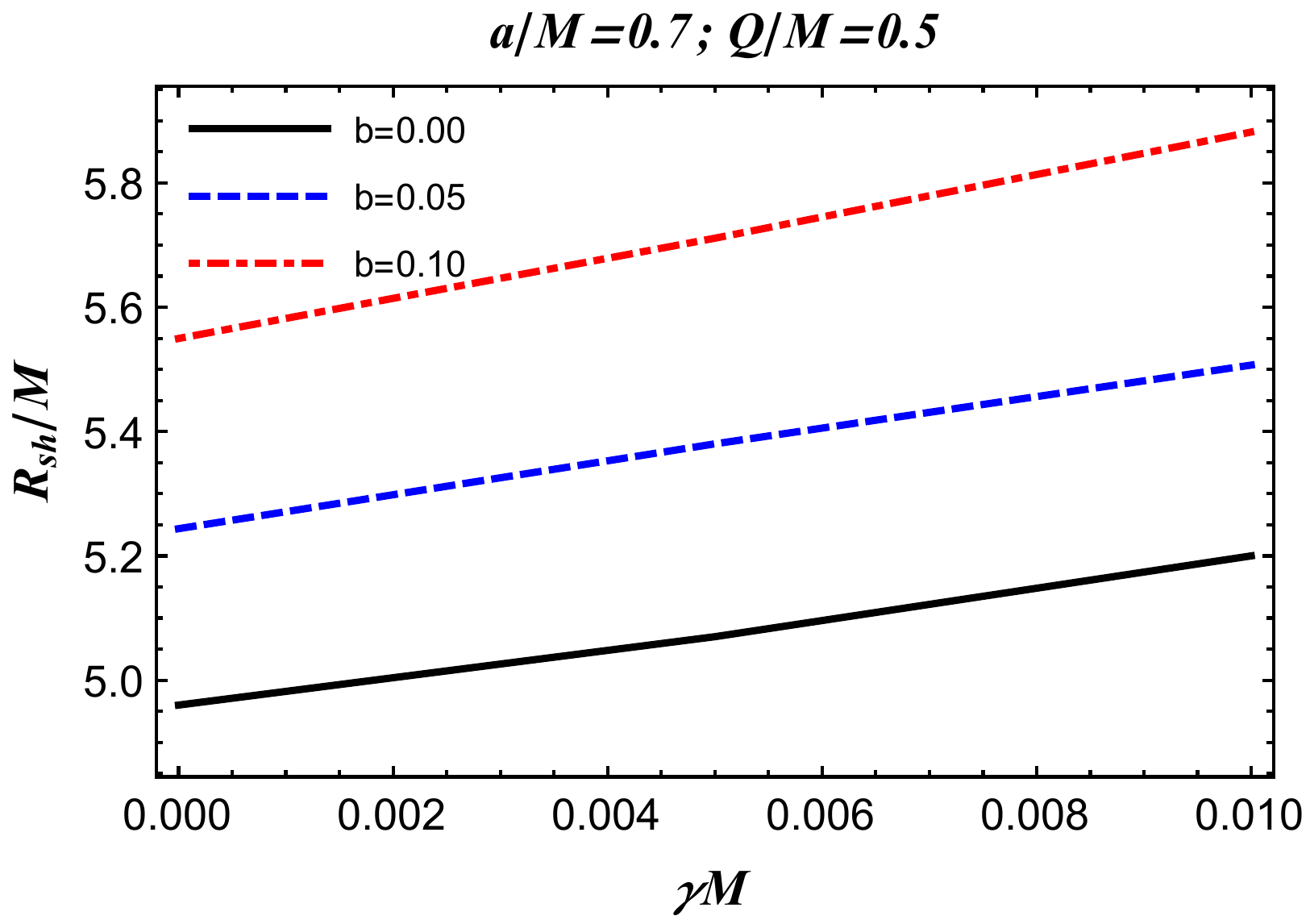}
  \end{center}
\caption{Radius of shadow. }\label{plot:radishadow}
\end{figure*}

\begin{figure*}
 \begin{center}
   \includegraphics[scale=0.5]{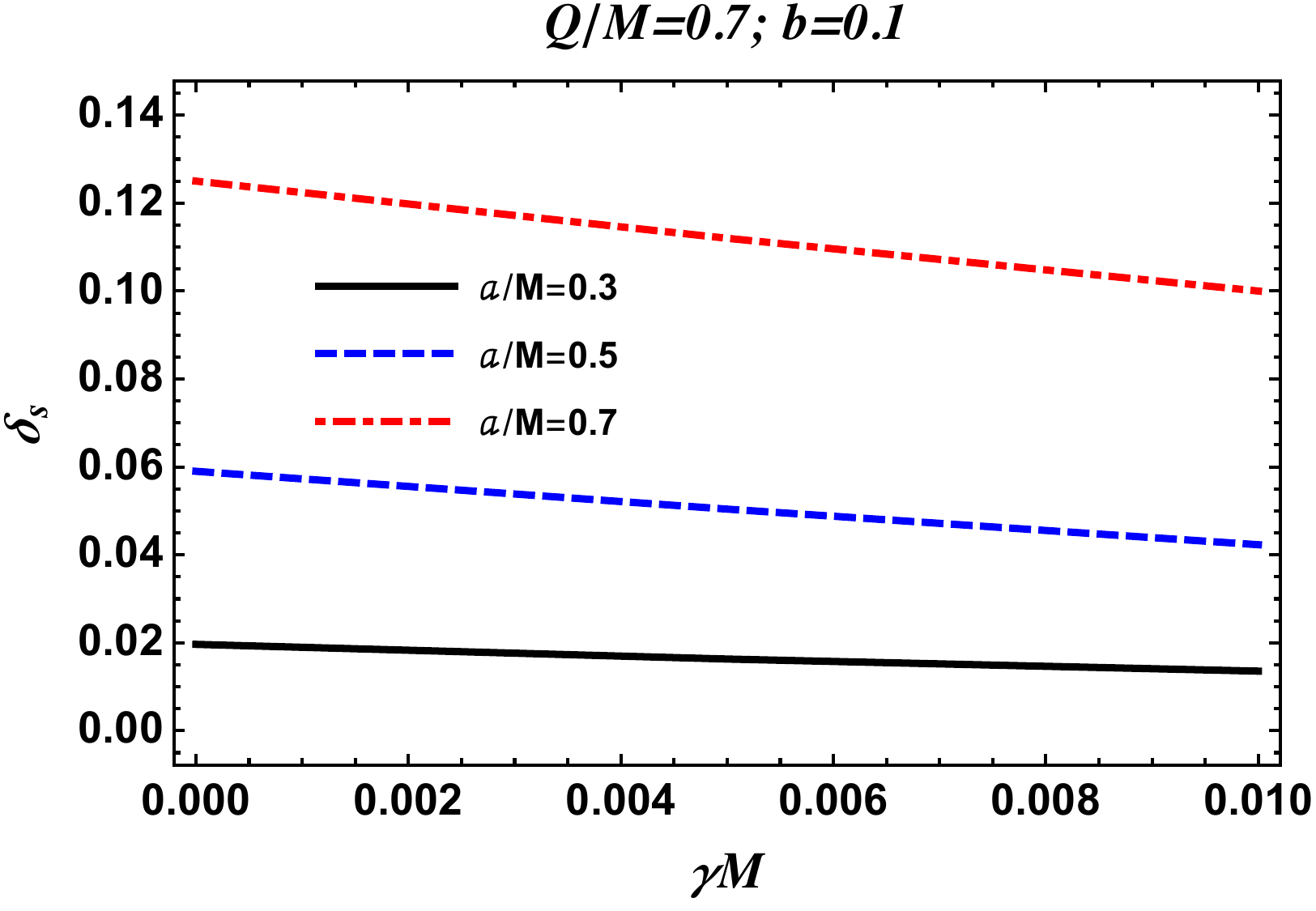}
   \includegraphics[scale=0.5]{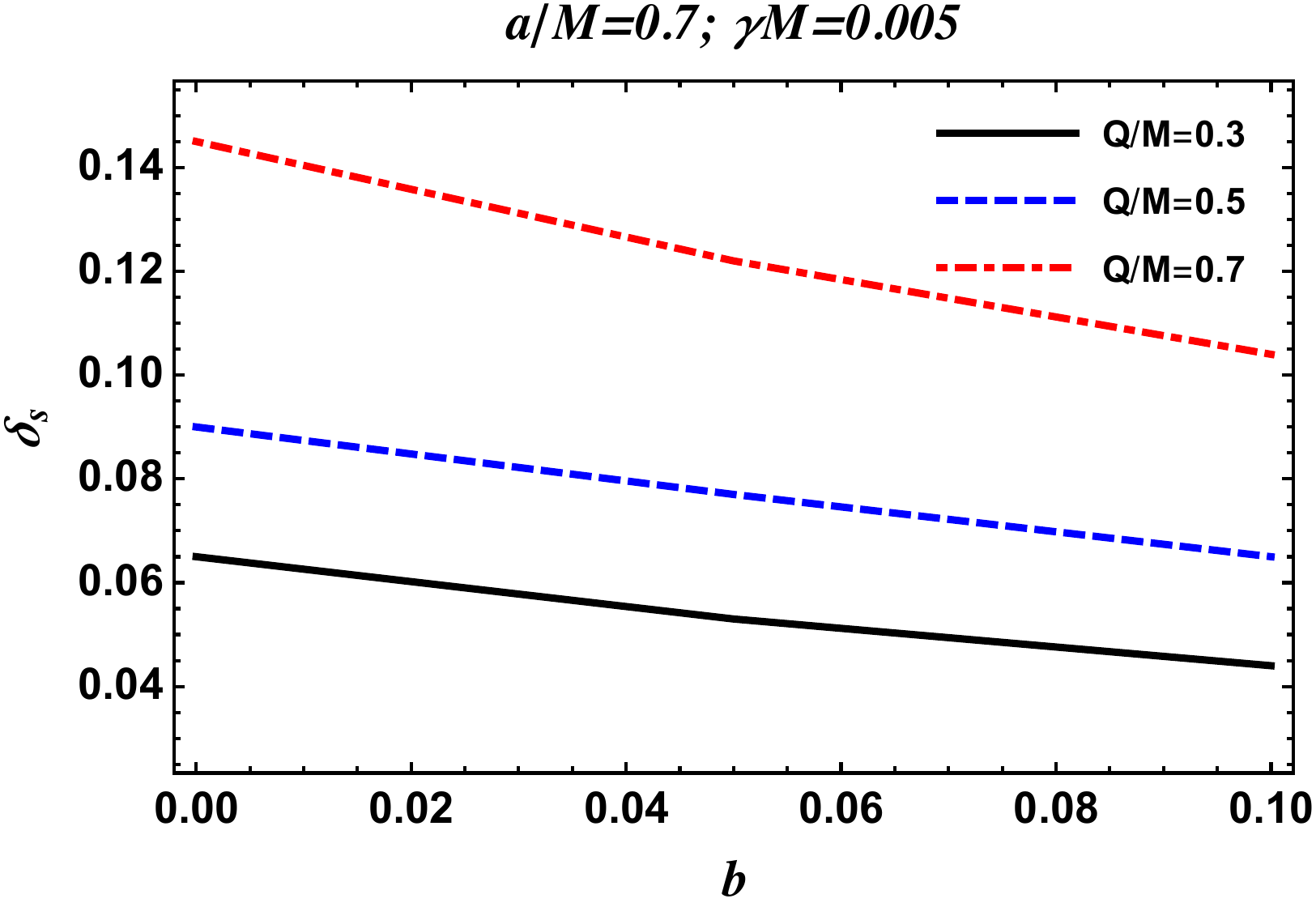}
  \end{center}
\caption{Distortion.}\label{plot:distortion}
\end{figure*}

Here we denote the radius of the circular shadow of the black hole by $R_{sh}$ and the distortion parameter $\delta_s$ is given by
\begin{equation}
\delta_{s}=\frac{D_{cs}}{R_{sh}},
\end{equation}
here $D_cs$ denotes the difference between the right endpoints of
the shadow cast by the black hole. The other observable $R_sh$ is given as
\begin{equation}
R_{sh} = \frac{(\alpha_t - \alpha_r)^2 + \beta_t^2}{2(\alpha_t - \alpha_r)}.
\end{equation}
In the Figs.~\ref{plot:radishadow} and \ref{plot:distortion}, we plot the observable $R_{sh}$ and $\delta_s$ as a function of the parameters $b$ and $\gamma$ and charge $Q$. We see that the observable $R_{sh}$ increase with the parameters $b$ and $\gamma$ and hence the size of the shadow cast by the KNKL black hole increases. While the effect of the charge $Q$ on the size of the shadow of the KNKL black hole is just opposite to that of the parameters $b$ and $\gamma$. From the Fig. \ref{plot:distortion} we notice that the observable $\delta_s$ which is responsible for the distortion in the shape of the shadow cast by the black hole increases with both $Q$ and $a$. While it decreases with the parameters $b$ and $\gamma$. This observation may constrain both the charge and spin of the black hole.

\subsection{Constraints on the parameters $b$ and $\gamma$ from the data provided by the EHT for M87* and SgrA*}

With the assumption that the supermassive black holes at the centre of the galaxies M87 and Milky way are surrounded by the CS and quintessence, we obtain upper limits on the quintessence parameters $\gamma$ and the SC parameter $b$, using the data provided by the EHT collaboration for the two supermassive black holes M87* and SgrA*. The angular diameter of the black hole shadow, the distance of the black hole from earth and the estimated mass of the black hole at the centre of the galaxy M87*, are given as $\theta_\text{M87*} = 42 \pm 3 \:\mu$as,  $D = 16.8$ Mpc and $M_\text{M87*} = 6.5 \pm 0.90$x$10^9 \: M_\odot$, respectively \cite{2}. For the supermassive black hole Sgr A* at the centre of the Milky way galaxy, the data obtained by the EHT collaboration is as $\theta_\text{Sgr A*} = 48.7 \pm 7 \:\mu$ , $D = 8277\pm33$ pc and $M_\text{Sgr A*} = 4.3 \pm 0.013$x$10^6 \: M_\odot$ (VLTI) \cite{Akiyama2022sgr}. Utilising the data provided by the EHT collaboration, we explore the diameter of the shadow of the black hole, per unit mass using the expression given as \cite{bambi2019},
\begin{equation}
    d_\text{sh} = \frac{D \theta}{M}\,.
\end{equation}
The diameter of the shadow can then be obtained from the expression $d_\text{sh}^\text{theo} = 2R_\text{sh}$. Thereby, the diameter of the image of the shadow cast by the black hole is $d^\text{M87*}_\text{sh} = (11 \pm 1.5)M$ for the supermassive black hole M87* and  $d^\text{Sgr A*}_\text{sh} = (9.5 \pm 1.4)M$ for the supermassive black hole Sgr A* at the cetre of the Milky way galaxy. taking in account the data released by the EHT collaboration, we have the constraints on the parameters $b$ and $\gamma$ for the supermassive black holes, one at the centre of the galaxy M87 and the other at the centre of the Milky way galaxy. For fixed values of the spin $a$ and charge $Q$ of the KNKL black hole, we show our results in the Tables~\ref{tab1}. We observe that the upper limit of the CS parameter $b$ increases when the quintessence parameter $\gamma$ decreases. This study of constraining the two parameters $b$ and $\gamma$, for the KNKL black hole, suggests that the effect of the SC may be stronger then the effect of the quintessence on the spacetime geometry of the KNKL black hole.

\begin{widetext}
\begin{center}
\begin{table}
\caption{\label{tab1}{The upper values of $\gamma$ and $a$ are tabulated for the supermassive BHs in the galaxy M87* and the Sgr A*. Note that we set $M=1$.}}
\begin{tabular}{|c|c|c|c|c|c|c|c|c|}
\hline
    & \multicolumn{2}{|c|}{Estimated values of parameters for M87* BH } \\
\hline
Parameters               &$Q=0.0$ and $a=0.5$                                                  &$Q=0.5$ and $a=0.5$   \\
\hline
$b$                      & $0.0000$\;\;\;\;\;$0.0200$\;\;\;\;\;$0.0400$          & $0.0000$\;\;\;\;\;$0.0200$\;\;\;\;\;$0.0400$                \\
$\gamma$                 & $0.0358$\;\;\;\;\;$0.0297$\;\;\;\;\;$0.0235$          & $0.0435$\;\;\;\;\;$0.0386$\;\;\;\;\;$0.0339$               \\
 \hline
     & \multicolumn{2}{|c|}{Estimated values of parameters for Sgr A* BH } \\
\hline
Parameters               &$Q=0.0$ and $a=0.5$                                                  &$Q=0.5$ and $a=0.5$   \\
\hline
$b$                      & $0.0000$\;\;\;\;\;$0.0200$\;\;\;\;\;$0.0345$          & $0.0000$\;\;\;\;\;$0.0200$\;\;\;\;\;$0.0400$              \\
$\gamma$                 & $0.0110$\;\;\;\;\;$0.0050$\;\;\;\;\;$0.0000$      & $0.0201$\;\;\;\;\;$0.0154$\;\;\;\;\;$0.0104$ \\
 \hline
\end{tabular}
\end{table}
\end{center}
\end{widetext}


\section{Rate of emission energy} \label{Sec:emission}

Due to the quantum fluctuations in a black hole spacetime the creation and annihilation of pairs of particles take place in the vicinity of the horizon of the black hole. During this process particles having positive energy escape from the black hole through the quantum tunneling. In the region where the Hawking-radiation takes place the black hole evaporates in a definite period of time. Here in this section we study the associated rate of the energy emission. Near a limiting constant value $\sigma_{lim}$, at a high energy the cross section of absorption of a black hole modulates slightly. As a consequence the shadow cast by the black hole causes the high energy cross section of absorption by the black hole for the observer located at infinity. The limiting constant value $\sigma_{lim}$, which is related to the radius of the photon sphere is given as \cite{21,24}
\begin{equation}\label{sigmalast}
\sigma_{lim} \approx \pi R_{sh}^2,
\end{equation}
where $R_{sh}$ denotes the radius of the black hole shadow,. The expression for the rate of the energy emission of a black hole is  \cite{21,24}
\begin{equation}\label{emissionenergyeq}
\frac{d^2 {\cal E}}{d\omega dt}= \frac{2 \pi^2 \sigma_{lim}}{\exp[{\omega/T}]-1}\omega^3,
\end{equation}
where $T=\kappa/2 \pi$ is the expression for the Hawking temperature and $\kappa$ is the notation used for the surface gravity. Combining eq.(\ref{sigmalast}) with the eq.(\ref{emissionenergyeq}), we arrive at an alternate form for the expression of emission energy rate as 

\begin{equation}\label{emissionenergyeqlast}
\frac{d^2 {\cal E}}{d\omega dt}= \frac{2\pi^3 R_{sh}^2}{e^{\omega/T}-1}\omega^3.
\end{equation}

\begin{figure*}
 \begin{center}
   \includegraphics[scale=0.65]{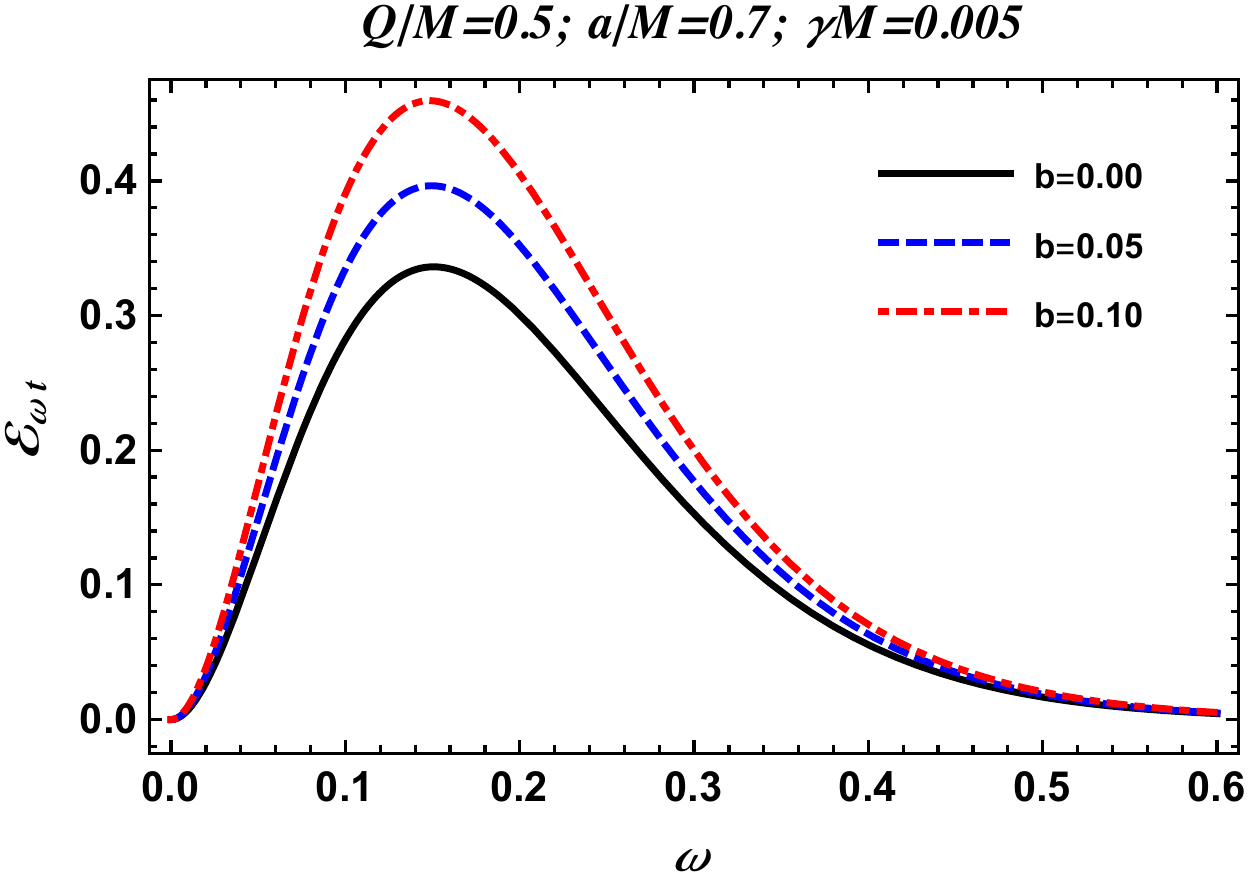}
   \includegraphics[scale=0.65]{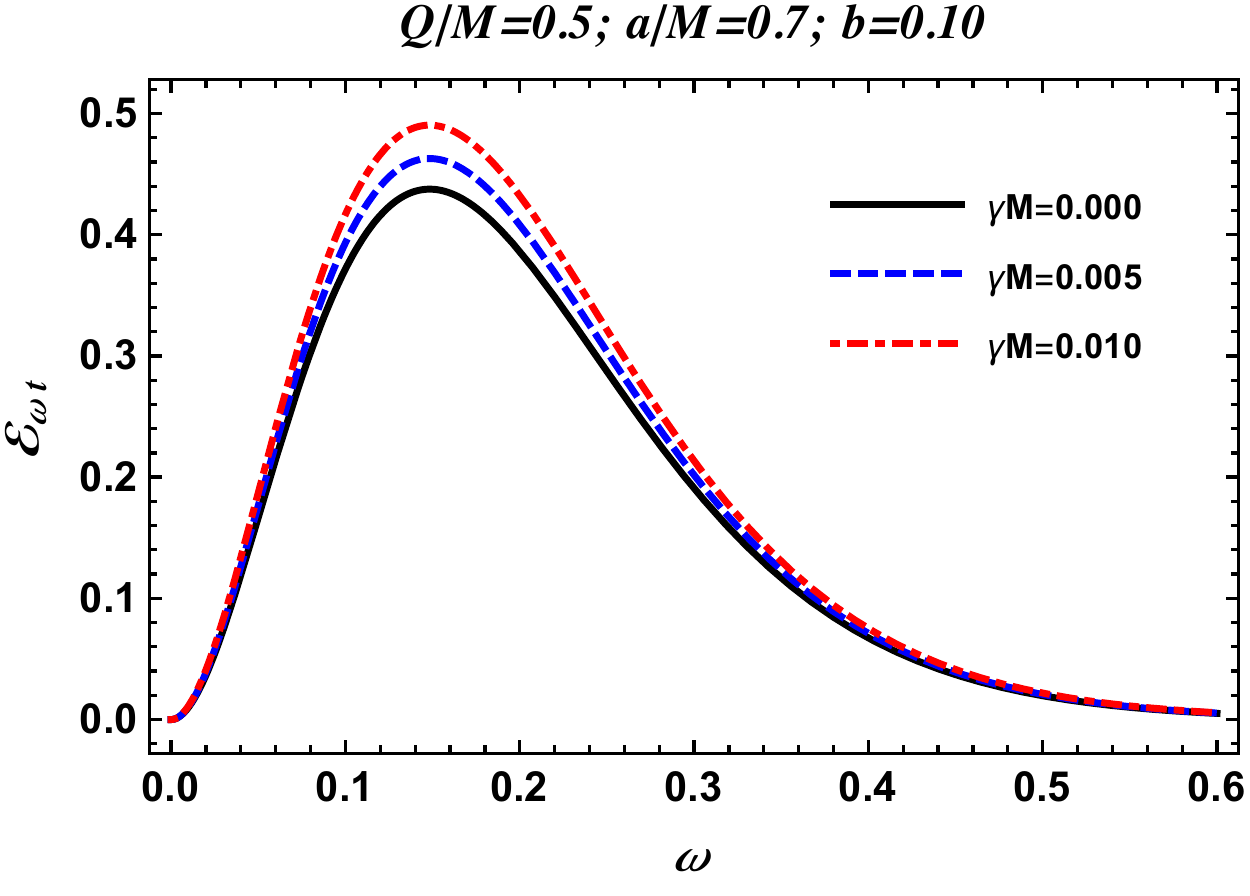}
  \end{center}
\caption{Emission energy.}\label{plot:emission}
\end{figure*}

Variation of the energy emission rate with respect to the frequency of photon $\omega$, for fixed values of the spin $a$ and charge $Q$ of the KNKL black hole and for different values of the parameters $b$ and $\gamma$ is represented in Fig.~\ref{plot:emission}. We see that with an increase in the values of the parameters $b$ and $\gamma$, the peak of the graph of the rate of the energy emission increases. This indicates that for the lower energy emission rate the evaporation process of the black hole is slow. It should be mentioned that for convenience we denote ${\cal E}_{\omega t}$ as $\frac{d^2 {\cal E}}{d\omega dt}$ in the Fig.~\ref{plot:emission}.

\section{Equatorial and polar QNMs and their relation with typical shadow radius} \label{Sec:QNMs}

 In this section we explore the relation between the typical shadow radius and QNMs. This correspondence is interesting and can be tested in the near future. In particular we know that QNMs are modes related to ringdown phase of the black hole and can be given in terms of the real and imaginary part as follows  $\omega=\omega_{\Re}- i \omega_{\Im}$ while the shadow radius is relevant for testing GR in the strong gravity regime using supermassive black holes. It has been shown that for asymptotically flat and static metrics there is a relation between angular velocity and real part of QNMs \cite{Cardoso:2008bp}. In particular one can then easily relate the shadow radius and the real part of QNMs in terms of the relation \cite{Jusufi:2020dhz,Cuadros-Melgar:2020kqn} 
	\begin{equation}
		R_{sh}=\frac{l+\frac{1}{2}}{\omega_{\Re}}.
	\end{equation}
	Note, however, that this relation can be violated in in modified theories of gravity like  Einstein–Lovelock theory, shown in \cite{Konoplya:2017wot}. Moreover in Ref. \cite{Yang:2021zqy} it was shown that the QNM frequency in the Eikonal limit reads
	\begin{eqnarray}
		\omega_{QNM}=(l+\frac{1}{2})\Omega_R-i \gamma_L \left(n+\frac{1}{2}\right)
	\end{eqnarray}
	where
	\begin{eqnarray}
		\Omega_R=\Omega_{\theta}+\frac{m}{l+\frac{1}{2}}\Omega_{prec},
	\end{eqnarray}
	in which $\Omega_{\theta}$ gives the orbital frequency in the polar direction. Moreover $\Omega_{prec}$ is known as the Lense-Thiring precision frequency of the orbit plane and $\gamma_L$ is the Lyapunov exponent of the orbit. It is well known that apart from the mass of the black hole (which is proportional to the shadow radius), due to the black hole spin the shadow gets distorted and the apparent shape of the shadow depends on the viewing angle. Therefore it is not possible in general to find a closed form or analytical expression for the shadow radius. 
	
	\begin{itemize}
		\item Case I: Viewing angle $\theta_0=\pi/2 $
	\end{itemize}
	Let us consider here the special case with the viewing angle $\theta_0=\pi/2 $. Moreover, we consider equatorial orbit which can be used to compute the typical shadow radius. To do so, we use the fact that the Lense-Thiring precision frequency is related to the orbital frequency and Keplerian frequency via
	\begin{eqnarray}
		\Omega_{prec}=\pm \Omega_{\phi}\mp \Omega_{\theta}\label{lt}
	\end{eqnarray}
	with
	\begin{eqnarray}\label{Omegaf}
		\Omega_{\phi}=\frac{-\partial _r g_{t \phi }\pm \sqrt{\left(\partial _r g_{t \phi }\right)^2-(\partial _r g_{t t})( \partial _r g_{\phi  \phi })}}{\partial _r g_{\phi
				\phi }}.
	\end{eqnarray}
	
	One can apply the same approach as the WKB analysis for Kerr quasinormal modes, then by writing $2 \int_{\theta_-}^{\theta_+} \sqrt{\Theta}\,d\theta=2\pi (L-|L_z|)$, which physically can be viewed as the Bohr-Sommerfeld Condition and can be compare with the eigenvalue problem in the $\theta$ direction for the Kerr quasinormal modes (see for details in Ref. \cite{Yang:2021zqy}). It was argued that
	\begin{eqnarray}
		\mathcal{K}+L_z^2 \simeq L^2-\frac{a^2 E^2}{2}\left(1-\frac{L_z^2}{L^2}\right),
	\end{eqnarray}
	and now if we divide the last equation by $E^2$,  we can obtain 
	\begin{eqnarray}
		\eta+\xi^2 \simeq \frac{L^2}{E^2}-\frac{a^2 }{2}\left(1-\frac{L_z^2}{L^2}\right).\label{xi1}
	\end{eqnarray}
	
	At this point we can make use of the correspondence \cite{Yang:2021zqy} 
	\begin{eqnarray}
		L_z  & \Longleftrightarrow & m\\ 
		E & \Longleftrightarrow & \omega_{\Re}\\
		L & \Longleftrightarrow & l+\frac{1}{2}
	\end{eqnarray}
	where one has $\omega_{\Re}=L \Omega_R$. In the limit $m=l>>1$, also known as the Eikonal limit, we have $\mu=m/(l+1/2)=1$ with
	\begin{eqnarray}
		\Omega_{prec}=\Omega_{\phi}-\Omega_{\theta},
	\end{eqnarray}
	one can then ind
	\begin{eqnarray}
		\Omega_R=\Omega_{\theta}+\Omega_{prec}=\Omega_{\phi}.
	\end{eqnarray}
	 In other words, these QNMs are related to the Kepler frequency which can be written as  \cite{Jusufi:2022tcw}
	\begin{eqnarray}
		\omega^{\pm}_{\Re}=(l+\frac{1}{2})\frac{-\partial _r g_{t \phi }\pm \sqrt{\left(\partial _r g_{t \phi }\right)^2-\partial _r g_{t t} \partial _r g_{\phi  \phi }}}{\partial _r g_{\phi
				\phi }},\label{rpart}
	\end{eqnarray}
	provided $m=l>>1$. We use the following definition to specify the typical shadow radius 
	\begin{eqnarray}
		\bar{R}_{sh}:=\frac{1}{2}\left(\alpha^+(r_{ph}^+)-\alpha^-(r_{ph}^-)\right),\label{de1}
	\end{eqnarray}
	where $\alpha^{\pm}(r_{ph})=\pm \sqrt{f(r_0)} \,\xi$ and $\eta(r_{ph}^{\pm})=0$. Now if we use Eq. \eqref{xi1} it follows that
	\begin{eqnarray}
		\xi^{\pm}=\pm \sqrt{\frac{(l+\frac{1}{2})^2}{\omega^2_{\Re}(r_{ph}^{\pm})}-\frac{a^2}{2}(1-\mu^2)}.
	\end{eqnarray}
	
	Combining these equation we arrive at \cite{Jusufi:2022tcw}
	\begin{eqnarray}\notag
		\bar{R}_{sh}&=&\frac{\sqrt{f(r_0)}}{2}\sqrt{\frac{(l+\frac{1}{2})^2}{\omega^2_{\Re}(r_{ph}^+)}-\frac{a^2}{2}(1-\mu^2)}\\
		&+& \frac{\sqrt{f(r_0)}}{2}\sqrt{\frac{(l+\frac{1}{2})^2}{\omega^2_{\Re}(r_{ph}^-)}-\frac{a^2}{2}(1-\mu^2)}.
	\end{eqnarray}
	where 
	\begin{eqnarray}
	    f(r_0)=1-b-\frac{2M}{r}+\frac{Q^2}{r^2}-\gamma r^{-3\omega_{q}-1}|_{r_0}.
	\end{eqnarray}
	and $r_0$ is the location of the observer. In other words due to the CS parameter the spacetime topology is globally conical and not asymptotically flat hence the shadow radius measured by the distant observer is modified. The correspondence is precise if we consider the eikonal limit, that is, if we set $\mu=\pm 1$ (namely $ [(m=\pm l)]$, yielding 
	\begin{equation}
		\bar{R}_{sh}(\mu=\pm 1)=\left(l+\frac{1}{2}\right)\frac{\sqrt{f(r_0)}}{2}\left(\frac{1}{\omega_{\Re}(r_{ph}^+)}-\frac{1}{\omega_{\Re}(r_{ph}^-)}\right).
	\end{equation}

	We can obtain the static case when $\omega^+_{\Re}=-\omega^-_{\Re}=\omega_{\Re}$ yielding 
		\begin{equation}
		\bar{R}_{sh}=\sqrt{f(r_0)}\frac{l+\frac{1}{2}}{\omega_{\Re}}.
	\end{equation}
	 We see that for asymptotically flat spacetime $f(r_0)\to 1$ and the last equation reduces to Eq. (27). 
	Using the metric functions and after some algebraic manipulation in the Eikonal limit we can use Eq. \eqref{rpart} which can be further simplified as follows \cite{Jusufi:2022tcw}
	\begin{eqnarray}
		\omega_{\Re}^{\pm}=(l+\frac{1}{2}) \frac{1}{a \pm \sqrt{\frac{ 2 r_{ph}^{\pm}}{f'(r)|_{r_{ph}^{\pm}}}} }.
	\end{eqnarray}
	We can rewrite the typical shadow radius equation, to obtain a simple equation
	\begin{equation}
		\bar{R}_{sh}=\frac{\sqrt{2 f(r_0)}}{2}\left(\sqrt{\frac{ r_{ph}^{+}}{f'(r)|_{r_{ph}^{+}}}}+\sqrt{\frac{ r_{ph}^{-}}{f'(r)|_{r_{ph}^{-}}}}\right).\label{rs}
	\end{equation}
	The last equation is nothing but the result which was obtained previously in Ref. \cite{Yang:2021zqy}, where  the points $r_{ph}^{\pm}$ were determined by solving \cite{Jusufi:2022tcw,Jusufi:2020dhz}
	\begin{equation}
		r_{ph}^2-\frac{2 r_{ph}}{f'(r)|_{r_{ph}^{\pm}}}f(r_{ph})\mp 2 a \sqrt{\frac{2 r_{ph}}{f'(r)|_{r_{ph}^{\pm}}}}=0.
	\end{equation}
	 
	In Table I we present the numerical values for the equatorial QNMs of the KNKL black hole for a given domain of parameters. With the increase of $l$ we normally expect the precision to increase. By means of Eq. (44) we can easily obtain the typical shadow radius using the values for QNMs. 
	\begin{table*}[tbp]
		\begin{tabular}{|l|l|l|l|}
			\hline
			\multicolumn{1}{|c|}{ } &  \multicolumn{2}{c|}{ Equatorial modes  } &   \multicolumn{1}{c|}{  Polar modes  }
			\\ 
			\hline
			$l$ & \quad $\omega_{\Re}^+$  & \quad $\omega_{\Re}^-$ & \quad $\omega_{\Re}$\\
			\hline
			1 & 0.394368035 & -0.250600581  &  0.306612155 \\
			2 & 0.657280058 & -0.417667636 &  0.511020260 \\
			3 & 0.920192081 & -0.584734690 &  0.715428363 \\
			4 & 1.183104105 & -0.751801745  &  0.919836467  \\
			5 & 1.446016128 & -0.918868799 &  1.124244572 \\
			6 & 1.708928152 & -1.085935854 &  1.328652675  \\
			7 & 1.971840175 & -1.253002909 &   1.533060779  \\
			8 & 2.234752199 & -1.420069964 &  1.737468883 \\
			9 & 2.497664222 & -1.587137018 &  1.941876987 \\
			10 & 2.760576245 & -1.754204073  &  2.146285091  \\
			\hline
		\end{tabular}
		\caption{ \label{table4} Numerical values of the real part of QNMs for equatorial modes and polar modes. We have set $M=1, Q/M=0.5, a/M=0.5, \gamma =0.001, \omega_q=-2/3$ and $b=0.001$.}\label{tab2}
	\end{table*}
	
	\begin{itemize}
		\item Case II: Viewing angle $\theta_0=0$ \&  $\theta_0=\pi$
	\end{itemize}
	Our second example is to consider the polar orbit $\theta=0$ along with the viewing angle for the observer: $\theta_0=0$ \&  $\theta_0=\pi$. For the polar orbit, we know that the azimuthal angular momentum is zero, i.e., $L_z=0$.  Using the circular geodesics i.e., $\dot{r}^2$, it follows that
	\begin{equation}
		(r^2+a^2)^2-[r^2f(r)+a^2]R_s^2=0,\label{def2}
	\end{equation}
	along with
	\begin{equation}
		4 r (r^2+a^2)-2 r f(r)R_s^2-r^2f'(r)R_s^2=0,
	\end{equation}
	in which we have the impact parameter $R_s^2=\mathcal{K}/E^2+a^2$ (see, \cite{Dolan:2010wr}).  Now by using Eq. \eqref{def2} we obtain 
	\begin{equation}
		R_s^{\pm}=\pm \frac{a^2+r^2}{\sqrt{r^2 f(r)+a^2}}|_{r_{ph}},
	\end{equation}
	For the typical shadow radius we assume the following definition $\bar{R}_{sh}:=(\sqrt{f(r_0)}R_s^+-\sqrt{f(r_0)}R_s^-)/2$, yielding 
	\begin{equation}
		\bar{R}_{sh}=\sqrt{f(r_0)}\frac{a^2+r^2}{\sqrt{r^2 f(r)+a^2}}|_{r_{ph}},\label{eq61}
	\end{equation}
	where $r_{ph}$ can be found by solving the relation
	\begin{eqnarray}
		(a^2+r_{ph}^2)^2-\frac{4 [r_{ph}^2 f(r_{ph})+a^2](a^2+r_{ph}^2)}{r_{ph} f'(r_{ph})+2 f(r_{ph})}=0.
	\end{eqnarray}
Using Eq. (32) and taking $L_z=0$, for the typical shadow radius we obtain 
	\begin{eqnarray}
		\bar{R}_{sh}=\sqrt{f(r_0)}\sqrt{\frac{(l+1/2)^2}{\omega^2_{\Re}(r_{ph})}+\frac{a^2}{2}}.
	\end{eqnarray}
   One can easily observe that for the nonrotating case the shadow radius reduces to Eq. (45) as we expect. On the other hand if we combine Eqs. (45) and (52) the real part of QNMs can be expressed as follows
	\begin{equation}
		\omega_{\Re}=(l+\frac{1}{2})\sqrt{\frac{2(r^2f(r)+a^2)}{2(a^2+r^2)^2-a^2(r^2f(r)+a^2)}}|_{r=r_{ph}}.
	\end{equation}
	
		Finally, in Table~\ref{tab2}, we have presented the numerical values for the polar QNMs. We expect that with the increase of $l$ the precision of the numerical values for the QNMs frequency increases.

\section{Shadow cast by the Kerr-Newman-Kiselev-Letelier black hole in the presence of plasma}\label{Sec:plasmashadow}

In GR mostly the influence of the medium on the light rays passing through it is neglected. Nonetheless, for instance the Solar corona influences the travel time of the radio signals and also their  angle of deflection very close to the Sun. This phenomena suggests the presence of a medium which can give us some nontrivial physics. Therefore, it is of interest to study the astrophysically relevant processes like black hole shadow in the presence of a plasma medium. Consequently in this section we investigate the shadow cast by the KNKL black hole in a plasma medium.

\begin{figure*}
 \begin{center}
   \includegraphics[scale=0.65]{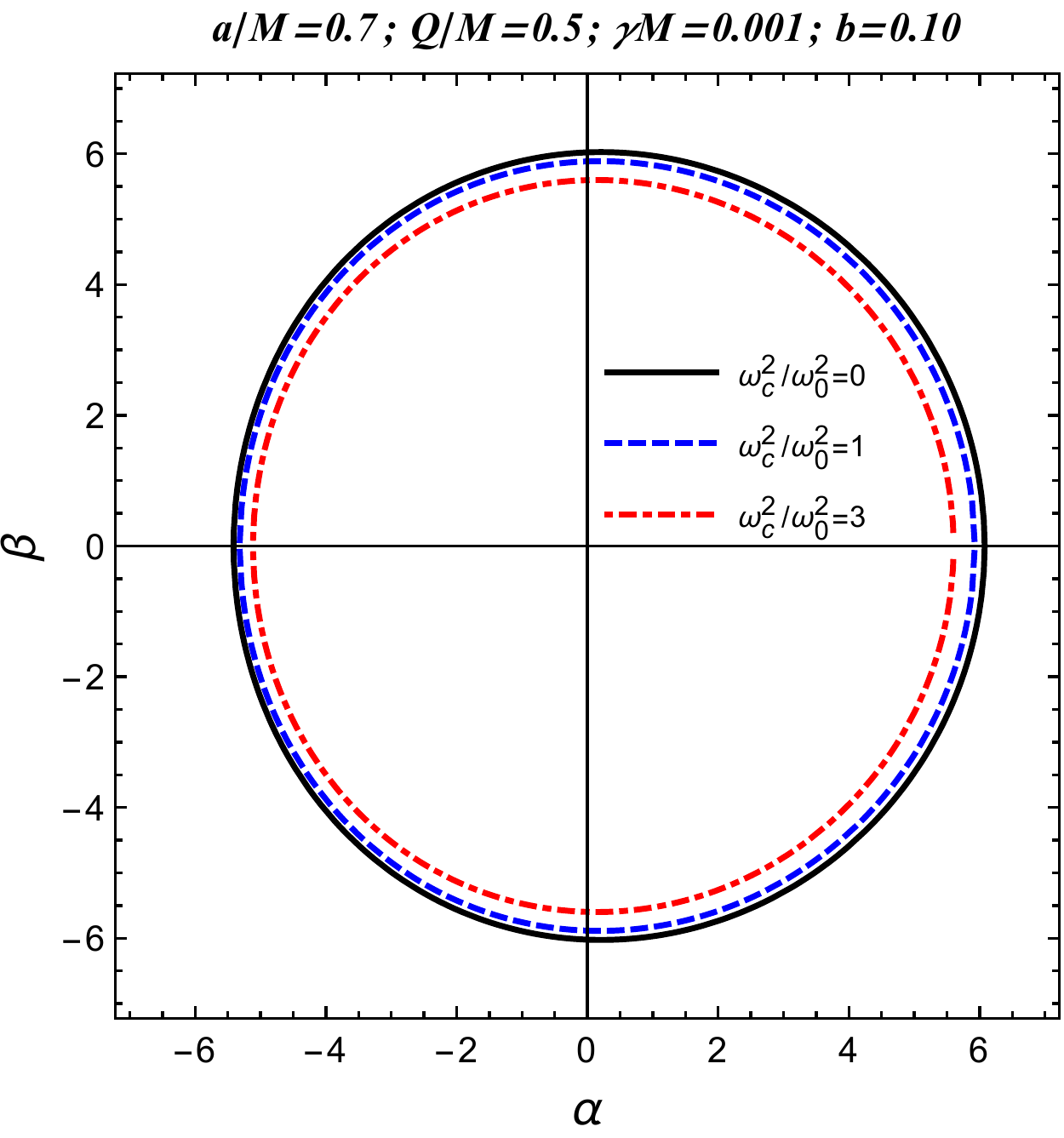}
   \includegraphics[scale=0.65]{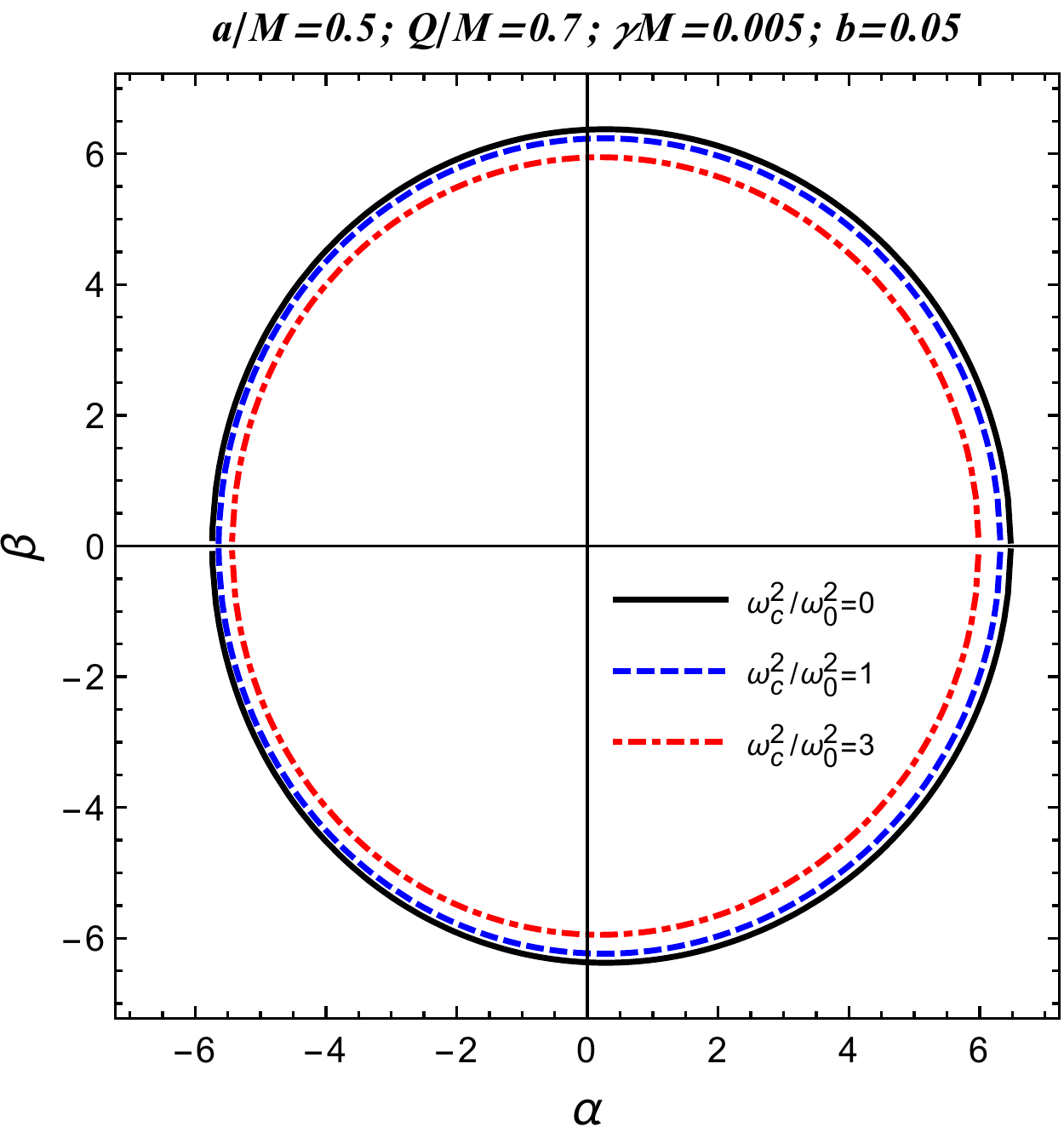}
   \includegraphics[scale=0.65]{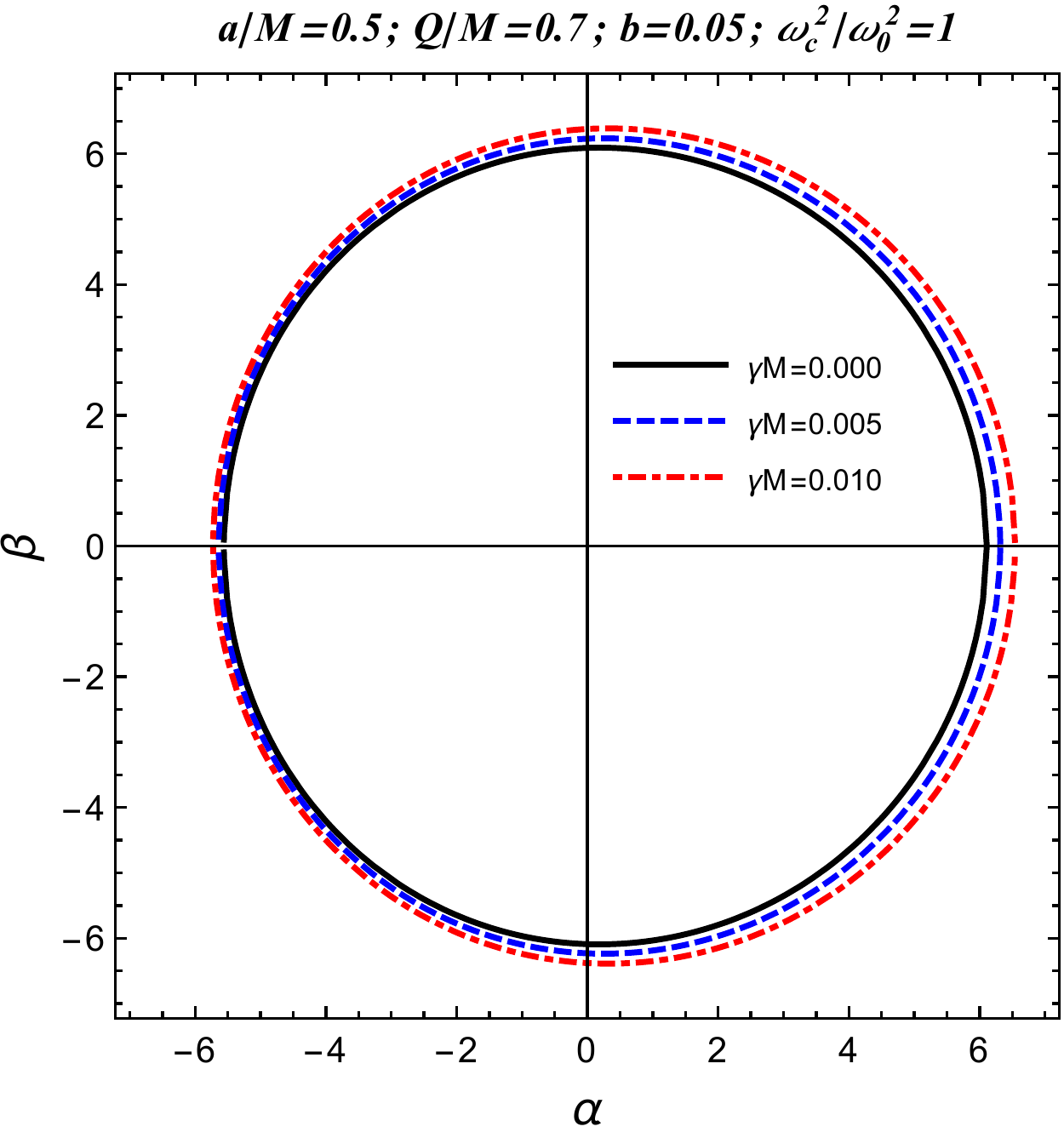}
   \includegraphics[scale=0.65]{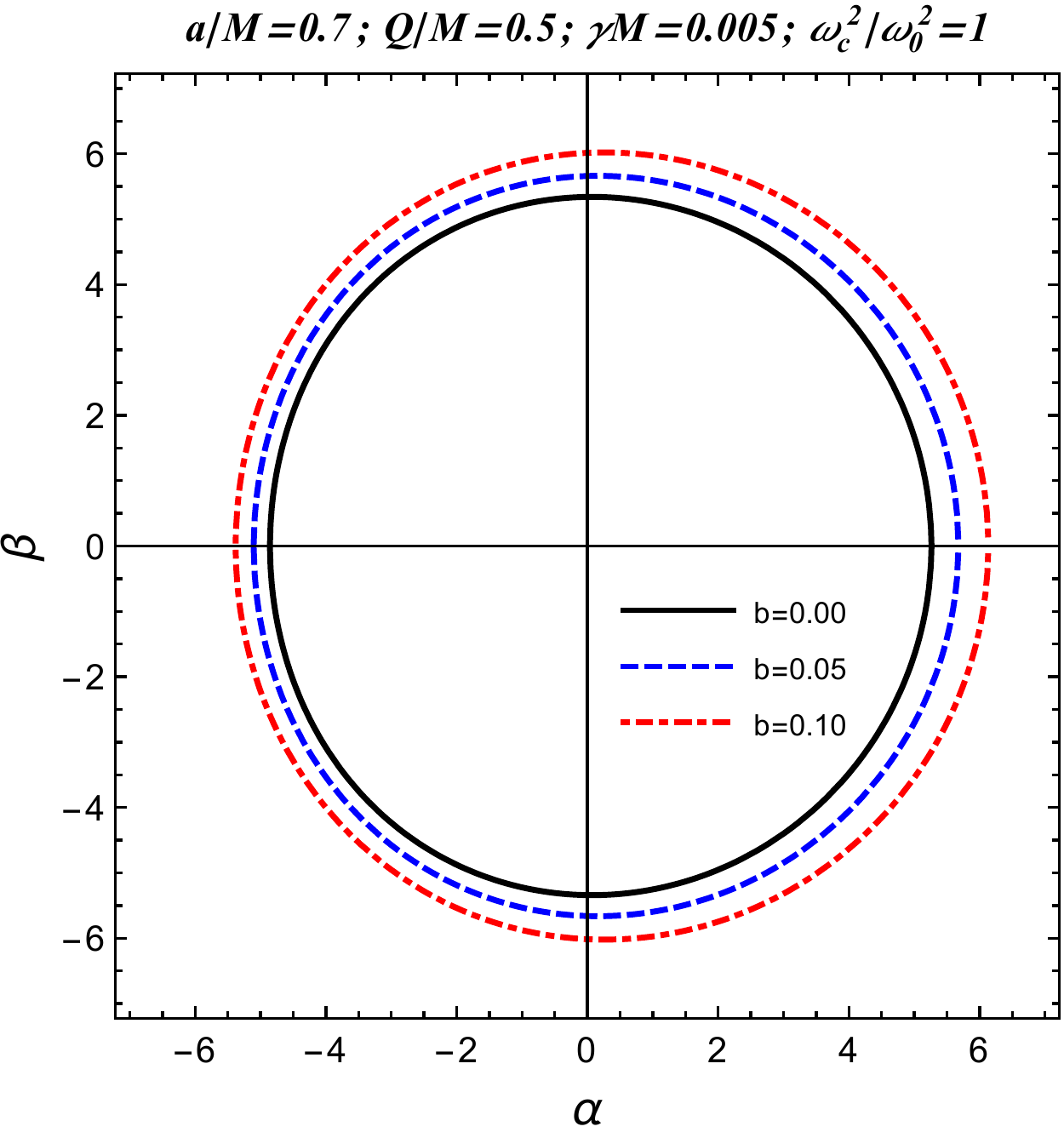}
  \end{center}
\caption{Shadow of the KNKL black hole in the presence of plasma. }\label{plot:shadowplasma}
\end{figure*}

\subsection{Dynamics of photon in a plasma medium}

Here we use the the Hamiltonian for the light ray moving around a black hole which is surrounded by plasma to obtain the equation of motion in the following form~\cite{Synge:1960b,Atamurotov217}:
\begin{equation}
\mathcal{H}=\frac{1}{2}\Big[g^{\mu\nu}p_{\mu}p_{\nu}+\omega_{\text{p}}(x)^2\Big]\ , \label{eq:hamiltonnon}
\end{equation}
where the electron plasma frequency $\omega_p$ is defined as~\cite{Bin:2010a}
\begin{equation}
     \omega_p(x)^2=\frac{4\pi e^2}{m_e}N_e(x)\ ,\label{rr1}
\end{equation}
with the charge and mass of electron as $e$ and $m_e$, respectively. The number density of electrons is $N_e$.   

In the case of photon the Hamilton-Jacobi equation is given as
\begin{equation}
    \mathcal{H}\Big(x,\frac{\partial S}{\partial x}\Big)=0\ . 
    \label{HJe}
\end{equation}
Here we adopt the  method of separation of variables and write the action in the following separable form 
\begin{equation}
    S=-\omega_0 t+p_{\phi}\phi+S_{r}(r)+S_{\theta}(\theta)\ ,
    \label{jaction}
\end{equation}
where the angular momentum and energy of the particle are $p_{\phi}$, $\omega_0$, which are the conserved quantities.
Recently, it is presented and discussed for the rotating BH~\cite{29}, that the plasma frequency can be written in the following form \cite{29,31}:
\begin{equation}
     \omega_p(x)^2=\frac{h(r)+g(\theta)}{\rho^2}\ ,
     \label{omega}
\end{equation}
where the functions $h(r)$ and $g(\theta)$ are related to the radial and the angular parts\cite{29}, respectively.
We use (\ref{jaction}) and (\ref{omega}) into Eq.~(\ref{HJe}) and
get the following expression
\begin{eqnarray}\label{eq:main}
    0=\frac{a^2\Delta \sin^2\theta-(r^2+a^2)^2}{\Delta}\omega_0^2 + 
    \frac{4aMre^{-l/r}}{\Delta}\omega_0p_\phi +\nonumber\\
    \Delta (S'_r)^2
    +(S'_\theta)^2+ 
    \frac{\Delta-a^2\sin^2\theta}{\Delta \sin^2\theta}p^2_\phi+
    h(r)+g(\theta)\ , 
\end{eqnarray}
in the KNKL black hole spacetime with a plasma medium. Next we  use the Carter constant ${\cal K}$ to separate the equation into the  the following two parts as\cite{31}
\begin{eqnarray}
    (S'_\theta)^2+\big(a\omega_0\sin\theta-\frac{p_\phi}{\sin\theta}\big)^2+g(\theta)=\mathcal{K}, \\
   \frac{1}{\Delta} -\Delta (S'_r)^2
    ((r^2+a^2)\omega_0-ap_\phi)^2-h(r)=\mathcal{K}\ . 
\end{eqnarray}
Using the Eq.~(\ref{eq:main}) we can get the equation of motion in the spacetime of the KNKL black hole with the plasma medium as\cite{31} 
\begin{eqnarray}
    \rho^{2}\frac{dt}{d\tau}&=&a(p_\phi-a\omega_0\sin^{2}\theta)
        +\frac{r^{2}+a^{2}}{\Delta} P(r),\label{rhot}\\
    \rho^{2}\frac{dr}{d\tau}&=&\pm\sqrt{\mathcal{R}},\label{Rad}\\
    \rho^{2}\frac{d\theta}{d\tau}&=&\pm\sqrt{\Theta},\\
    \rho^{2}\frac{d\phi}{d\tau}
     &=&\frac{p_\phi}{\sin^{2}\theta}-a\omega_0+\frac{a}{\Delta}P(r)\ , 
     \label{rhophi}
\end{eqnarray}
where $P(r)$ is given by%
\begin{equation}
    P(r)=(r^2+a^2)\omega_0-ap_\phi\ , 
\end{equation}
the functions $\mathcal{R}$ and $\Theta$ are related to the radial and angular equations of motion, respectively, and are expressed as
\begin{eqnarray}
\mathcal{R}&=&P(r)^2-\Delta\Big[\mathcal{Q}+(p_\phi-a\omega_0)^2+h(r)\Big],\\
\Theta&=&\mathcal{Q}+\cos^2\theta\left(a^2\omega^2-p_\phi^2\sin^{-2}\theta\right)-g(\theta),
\end{eqnarray} 
where $\mathcal{Q}=\mathcal{K}-(p_\phi-a\omega_0)^2$.

\subsection{Shadow of the Kerr-Newman-Kiselev-Letelier black hole in the presence of plasma}

To discuss the shadows of black holes we evaluate the boundary of the circular light rays. We use the conditions given as $\mathcal{R}=0=\mathcal{R}'$ to obtain the constants of motion in terms of the radius $r$ of the circular orbits of photons as \cite{31}
\begin{eqnarray}
    \mathcal{Q}&=&\frac{\left(a p_\phi - \omega_0 \left(a^2+r^2\right)\right)^2}{\Delta}-(p_\phi-a \omega_0)^2-h(r), \\
    p_\phi&=&\frac{\omega_0}{a}\left[r^2+a^2-
    \frac{\Delta}{a \Delta'}  \left(\sqrt{4 r^2-\frac{h'(r) \Delta'(r)}{\omega_0}}+2 r \right)\right].\nonumber \\ 
\end{eqnarray}

Now we use the celestial coordinates to present the silhouette of the shadow cast by black hole in the presence of plasma. It can be represented in the following form\cite{Kimetnoflat2020,31}
\begin{eqnarray}
    \alpha&=&-\frac{p_\phi}{\omega_0\sin{\theta_0}}\sqrt{f(r_0)},\label{eq:shadowplasma1}
\\
    \beta&=&\pm\frac{\sqrt{f(r_0)}}{\omega_0}\sqrt{\mathcal{Q}+\cos^2{\theta_0}\Big(a^2\omega_0^2-\frac{p_\phi^2}{\sin^2{\theta_0}}\Big)-g(\theta_0)}.\nonumber \\\label{eq:shadowplasma2}
\end{eqnarray} 
%

To consider the size and the shape, as functions of the spacetime parameters, we will consider the well-known case of the dust that is at rest at infinity and was first used by Shapiro \cite{Shapiro1974}, for the plasma in our current analysis. In the rotating spacetime the mass density, and by Eq. \ref{rr1}, the squared plasma frequency, go as $r^{-3/2}$, being independent of $\theta$ to a very good approximation\cite{31}. However, such a plasma distribution cannot be put into the separable form given by Eq. \ref{omega}. Therefore, following Ref. \cite{29,31}, we take the frequency to have an additional angular dependency by choosing as~\cite{31}:
\begin{equation}
    h(r)=\omega_c^2\sqrt{M^3r}, \label{eq:plasma1}
\end{equation} 
\begin{equation}
    g(\theta)=0, \label{eq:plasma2}
\end{equation} 
from that~\cite{31}
\begin{equation}
    \omega_p^2=\omega_c^2\frac{\sqrt{M^3r}}{r^2+a^2\cos^2\theta}\ , 
\end{equation}
where $\omega_c$ is a constant\cite{31} and $M$ represents the mass of the black hole.

We combine Eqs.~(\ref{eq:shadowplasma1}), (\ref{eq:shadowplasma2}), (\ref{eq:plasma1}) and (\ref{eq:plasma2}), in the equatorial plane to get the shadow cast by the KNKL black hole and the plots in the Fig.~\ref{plot:shadowplasma} show the shadow of the black hole for the different values of the spacetime and the plasma parameters. From the Fig.~\ref{plot:shadowplasma}, we observe that with an increase in the plasma parameter the size of the shadow of the KNKL black hole decreases, for fixed values of all other spacetime parameters. Further we see that the shadow of the KNKL black hole increase for the increasing values of the parameter $\gamma$ and $b$, if we keep the plasma parameter and the other spacetime parameters fixed. Hence the presence of plasma shrinks the shadow cast by the KNKL black hole. Further, the nature of both the parameters $\gamma$ and $b$ is still repulsive even in the presence of the plasma medium.   

\section{Conclusions and discussions}
\label{Sec:Conclusions}

In this study we have discussed the photon motion and the related phenomena of black hole shadow in the KNKL black hole spacetime. In particular we have investigated the influence of the CS parameter $b$ and the quintessence parameter $\gamma$ on the photon motion and the black hole shadow. In Fig.~\ref{plot:horizon} we have seen that the size of the horizon of the KNKL black hole increases for the values of the parameters $\gamma$ and $b$, for fixed values of the spin $a$ and charge $Q$, and therefore both the parameters $b$ and $\gamma$ are repulsive in nature. In the Fig.~\ref{plot:effpot} we have plotted the effective potential for the photon, where it is decreasing with the increasing values of the parameters $b$ and $\gamma$. For the increasing values of the parameters $b$ and $\gamma$, the unsuitability of the photon circular orbits also decreases. In the Fig.~\ref{plot:shadow}, We have seen that the size of the shadow cast by the KNKL black hole increase with increase in the values of the parameters $b$ and $\gamma$, for fixed values of the spin $a$ and charge $Q$ of the black hole. Hence the repulsive nature of the quintessence and CS is confirmed again, for the KNKL black hole spacetime. In the same Fig. we have also seen that the size of the KNKL black hole shadow decreases for the increasing values of the parameters $a$ and $Q$, when the values of the other parameters are kept constant. In the Fig.~\ref{plot:radishadow} the radius of the shadow cast by the KNKL black hole is plotted against the parameters $b$ and $\gamma$ where it increase with $b$ and also with $\gamma$ for different values of other spacetime parameter. We have represented the distortion parameter $\delta_s$ for different values of the spin parameter $a$ and charge $Q$ of the KNKL black hole. In the Fig.~\ref{plot:distortion} We have observed that the parameter $\delta_s$ decrease with $b$ and $\gamma$. In the same Fig. we have noticed that the distortion of the shadow cast by the KNKL black hole increases as the values of the parameters $a$ and $Q$ are increasing. It may be concluded that the shadow cast by a fast rotating and highly charged black hole would be more distorted. This observation may be helpful for estimating the values of the spin $a$ and the charge $Q$ of black holes. 

Using the date of the EHT collaboration for the angular diameter of the supermassive black hole shadow, the distance of the supermassive black hole from earth and the estimated mass of the supermassive black hole at the centre of the galaxy M87 and Milky way, We have calculated the upper limits on the CS parameter $b$ and quintessence parameter $\gamma$ in the case of the KNKL black hole. We have seen that the parameter $b$ increases when the quintessence parameter $\gamma$ decreases. This behaviour of the two parameters $b$ and $\gamma$, shows that the SC may have stronger effects on the the spacetime geometry of the KNKL black hole, as compared with the effects of the quintessence on it.

Further we have observed that for fixed values of the spin $a$ and charge $Q$ of the KNKL black hole, the rate of the emission energy increases for the increasing values of both the CS parameter $b$ and the quintessence parameter $\gamma$, against the frequency $\omega$ of the photon. Thus the quintessence and also CS accelerate the Hawking radiation process and therefore, the KNKL black hole may evaporate quicker than the Kerr-Newman black hole. Note here that by setting the CS parameter $b$ to $8\pi\eta^2$, where $\eta$ is the global monopole charge, the metric resembles the Kerr-Newman black hole with a global monopole charge. This scenario was studied recently in Ref. \cite{Haroon:2019new} with different equations of state: $\omega_q=-1/3$; $\omega_q=0$; and $\omega_q=1/3$. 

Using the correspondence between geometric-optics of the parameters of a QNMs and the conserved quantities along geodesics we found and analyzed the typical shadow radius and equatorial and polar QNMs. The typical shadow radius is affected by the CS parameter.

Increasing the value of the plasma parameter and Keeping all the other spacetime parameters constant, we have seen that the size of the shadow cast by the KNKL black hole decreases. This decrease in the size of the shadow cast by the black hole is due to the refraction of the electromagnetic radiation in the plasma medium around the KNKL black hole. Further, we have noticed that with the increasing values of the CS parameter $b$ and the quintessence parameter $\gamma$, in the presence of plasma the size of the shadow cast by the KNKL black hole increases for the fixed values of the other spacetime parameters and as well as the plasma parameter. Thus we have seen that the behaviour of the quintessence and also of the CS is still repulse even in the presence of the plasma medium.

\section*{Acknowledgements}
F.A. acknowledges the support of Inha University in Tashkent and research work has been supported by the Visitor Research Fellowship at Zhejiang Normal University. This research is partly supported by Research Grant FZ-20200929344 and F-FA-2021-510 of the Uzbekistan Ministry for Innovative Development. G. Mustafa is very thankful to Prof. Gao Xianlong from the Department of Physics, Zhejiang Normal University, for his kind support and help during this research. Further, G. Mustafa acknowledges the Grant No. ZC304022919 to support his Postdoctoral Fellowship at Zhejiang Normal University.
\bibliographystyle{apsrev4-1}
\bibliography{KNKreference}

\end{document}